\documentclass{article}

\usepackage[english]{babel}

\usepackage[letterpaper,top=2cm,bottom=2cm,left=3cm,right=3cm,marginparwidth=1.75cm]{geometry}

\usepackage{amsmath}
\usepackage{graphicx}
\usepackage[colorlinks=true, allcolors=blue]{hyperref}
\usepackage{authblk}

\makeatletter
\author[3,5]{M. Abbrescia}
\author[2,4]{S. Colafranceschi}
\author[3,5]{M. De Serio}
\author[1]{B. Liberti}
\author[2,7]{S. Meola}
\author[2]{A. Paoloni}
\author[3]{A. Pastore}
\author[2]{D. Piccolo}
\author[2,6]{G. Saviano}
\author[2,6]{C. Vendittozzi}

\renewcommand\AB@authnote[1]{%
  \ifnum#1=0\relax
  \else
    $^{#1}$%
  \fi
}
\makeatother
\author[0]{A. Rocchi}

\affil[1]{INFN, Sezione di Roma Tor Vergata, Via della Ricerca Scientifica 1, 00133 Roma, Italy}
\affil[2]{INFN, Laboratori Nazionali di Frascati, 00044 Frascati, Italy}
\affil[3]{INFN Sezione di Bari, Via E. Orabona 4, 70125 Bari, Italy}
\affil[4]{James Madison University (JMU), Harrisonburg, VA 22807, USA}
\affil[5]{Università degli studi di Bari, Dipartimento Interateneo di Fisica, Via Amendola 173, 70125 Bari, Italy}
\affil[6]{Sapienza Università di Roma, Dipartimento di Ingegneria Chimica Materiali Ambiente, Via Eudossiana 18, 00184 Roma, Italy}
\affil[7]{Università degli Studi Guglielmo Marconi, 00193 Roma, Italy}
\date{} % opzionale, rimuove la data

\title{A standalone simulation program for Resistive Cylindrical Chamber (RCC)}
\begin{document}
\maketitle

\begin{abstract}
In recent years, the Resistive Cylindrical Chamber (RCC) has been introduced as a novel gaseous detector, extending the well-established Resistive Plate Chambers (RPCs) to the case of cylindrical electrode geometry. Preliminary experimental studies have highlighted several promising features of this configuration, motivating the need for further systematic investigations of its operation. In contrast, from the simulation perspective, detailed studies of the RCC have not  been performed yet, despite the fact that the cylindrical geometry introduces new degrees of freedom-such as cylinder electrodes radii and voltage polarity- which lead to asymmetric behaviour of the avalanche development according to the polarity of the applied voltage between the electrodes.

In this work we present a standalone simulation program specifically designed to model avalanche growth and signal induction in both RPC and RCC geometries. The code implements a stepwise transport model for electron multiplication, includes approximate space-charge effects, and evaluates the induced signals on an external electrode. 

The simulation has been validated against experimental data for planar RPCs and subsequently applied to RCC geometries. The results demonstrate that key observables such as induced charge and efficiency for the planar geometry are well reproduced and highlights the role of electric-field asymmetry in the cylindrical configuration. These findings provide quantitative insights into the impact of detector geometry on avalanche dynamics. 

\end{abstract}

\section{Introduction}

In recent years, a novel gaseous detector, the Resistive Cylindrical Chamber \cite{Cardarelli2021}, was introduced as an extension of the well-established Resistive Plate Chambers \cite{Santonico1981} to the case of cylindrical electrode geometry.  
It consists of two coaxial cylindrical electrodes made of resistive material, separated by a thin gas gap. 
A high voltage is applied between the inner and outer electrodes, creating a radial electric field that allows the development of avalanches when ionizing particles cross the gas. 
The use of resistive electrodes ensures stable operation by limiting the discharge current, while the cylindrical geometry introduces new degrees of freedom, such as the electrode radii and the voltage polarity, making the device a particularly attractive subject of study.

Preliminary experimental tests, performed between 2022 and 2024 \cite{CardarelliRocchi2024}\cite{Rocchi2024}, have highlighted several promising features of this detector concept, which call for systematic studies in order to fully characterize its performance and to explore both its potential advantages and limitations.  

On the other hand, from the simulation perspective, the RCC has not been investigated in detail yet. 

The aim of this work is to develop a standalone simulation code for the study of Resistive Cylindrical Chambers and to compare their performance with those of conventional Resistive Plate Chambers. 
The emphasis is on providing a fast yet sufficiently accurate tool, enabling fast parameter scans and design studies without the computational overhead of full-scale microscopic simulations. 
The code is designed to implement a stepwise transport and avalanche-growth model in the gas, and includes space-charge effects, realistic electrode properties, and induced-charge calculations. 
Throughout the paper, we explicitly point out the limitations and approximations inherent to the adopted approach.

This work is realised in the framework of the \textit{TANGO\_RD} project (Toward A New Generation Of RPC Detectors), funded by INFN, whose goal is to study the performance of the novel RCC geometry and to benchmark its operation through a combined program of experimental measurements and dedicated simulations.

Several simulation strategies for RPCs have been developed in the past, spanning from simplified semi-analytical models to full microscopic Monte Carlo simulations.

Earlier work \cite{Abbrescia1998,Abbrescia1999} presented analytical and semi-empirical avalanche models, 
validating them against measured RPC efficiency and charge spectra.
Among the simplified approaches, a one-dimensional step model where space-charge effects are included via analytical formulas were developed in ~\cite{Lippmann2004}, and extensions ~\cite{Riegler2002,Riegler2016}.
Layered weighting-field analytic models, often used in Garfield++ tutorials~\cite{GarfieldTutorials}, provide closed-form solutions for induced-charge calculations, 
and have been employed extensively for both planar and non-planar geometries.

In other directions, a more formal and detailed approaches have been implemented in Garfield++ ~\cite{garfieldpp-manual}. 
Here microscopic electron transport and ionization simulations are combined with detailed space-charge calculations, 
providing a highly accurate description of the detector response at the cost of significantly higher computational requirements.
Although the microscopic approach they adopt is both elegant and extremely detailed, its applicability is quite limited in the case of Resistive Plate Chambers and Resistive Cylindrical Chambers.  
In these detectors, the avalanches typically involve more than $10^{6}$ electrons, making a fully microscopic description practically unfeasible in terms of execution time.  
Moreover, the results obtained are not always in agreement with all experimental data.  

For these reasons, simplified or semi-phenomenological approaches, able to capture the essential features of avalanche development and signal induction without tracking each individual electron, become necessary for realistic studies of RPCs and RCCs.

The standalone code presented here adopts a  stepwise approach, including a simplified model of the space charge contributions and incorporates electrode effects and weighting-field evaluations for both RPC and RCC geometries.

In this article we present the results of the simulation program applied to both standard Resistive Plate Chambers (RPCs) and Resistive Cylindrical Chambers (RCCs) operating with the so-called \emph{standard gas mixture}, widely adopted in RPC applications, which is composed of 95.2\% tetrafluoroethane (C$_2$H$_2$F$_4$), 4.5\% isobutane (iC$_4$H$_{10}$), and 0.3\% sulfur hexafluoride (SF$_6$).  

By applying the simulation framework to RPC and RCC geometries in these conditions, we aim to highlight both the common features and the specific differences introduced by the cylindrical configuration.

The paper is organized as follows.  
In Sections~\ref{avaModel} and~\ref{alphaModel} we describe the avalanche generation model adopted in our code, including the contribution of space charge effects, and we outline the procedure used to evaluate the characteristic parameters of the process.  
Section~\ref{sec:RPCsim} presents the validation of the code for the planar geometry of a standard RPC with a 2~mm gas gap, while Section~\ref{RCCsim} extends the validation to the case of a RCC.  

In Section~\ref{inducedCharge} we introduce the framework employed to evaluate the induced charge on an external electrode, both in the planar and in the cylindrical geometry. 
Section~\ref{ionizationModel} discusses the simulation of the number of primary clusters and the secondary electrons, enabling the modeling of more realistic conditions corresponding to the passage of a charged particle through the detector.  
Sections~\ref{FullSim} merges all the ingredients described in the previous sections in a full simulation procedure with specific applications to the RPCs and RCCs with different geometries in sections~\ref{RPCFullSim} and~\ref{RCCFullSim} respectively.   

Finally, Section~\ref{conclusions} draws some preliminary conclusions on the operating features of the novel RCC detector. An Appendix is included, collecting the derivation of some of the analytical formulae used in the code.

\section{Avalanche Development Model}
\label{avaModel}

An avalanche in an electric field \( E \) starting from a point \( x_0 \) evolves according to the Townsend exponential growth law characterized by an effective Townsend coefficient \( \alpha_{\mathrm{eff}} \) defined as the difference between the Townsend coefficient $\alpha$ and the attachment coefficient $\eta$. Specifically, the number of electrons \( n \) at a distance \( x \) along the field grows as

\begin{equation}
n(x) = n_0 \, e^{\int_{x_0}^{x} \alpha_{eff}(E(x')) \, dx'}
\end{equation}
where \(\alpha_{\mathrm{eff}} = \alpha_{\mathrm{eff}}(E)\) is a function of the local electric field \(E\).

To account for non uniform electric fields as in the case for RCCs and in general for the space charge effect due to positive ions left behind along the path, the avalanche path is divided into \( N \) steps (typically on the order of 100-1000). At each step, the number of electrons and equivalently the number of positive ions generated, are calculated using the Townsend formula with the local electric field value at that step. This stepwise approach allows the electric field \( E \), and consequently the effective Townsend coefficient \( \alpha_{\mathrm{eff}} \), to be considered constant within each step, enabling \( \alpha_{\mathrm{eff}} \) to be factored out of the integral in the expression for \( n(x) \).

The positive ions are approximated as distributed uniformly over a disk of radius \( a \), where \( a \) is computed step-by-step according to the relation 
 \begin{equation}
a = 2 \cdot D_T \sqrt{x},
\label{eq:TransverseRadius}
\end{equation}
where \( D_T \) is the transverse diffusion coefficient and \( x \) the distance traveled by the avalanche. With this definition of the radius, the RMS of the uniform charge distribution over the disk coincides with that  of the expected bi-dimensional Gaussian distribution resulting from transverse diffusion.
For the standard gas mixture, a value \( D_T = 1.2 \times 10^{-2}~\text{cm}/\sqrt{\text{cm}} \), as obtained with Magboltz code, has been used in the simulation.
 
 At each step, the external electric field is evaluated, and the field contribution from all previous ion disks (space charge) is subtracted according to the following formula 
 \begin{equation}
E(s) = \frac{\sigma}{2 \epsilon_0} \left(1 - \frac{s}{\sqrt{s^2 + a^2}}\right),
\label{eq:SpaceChargeDisk}
\end{equation}
where $\sigma $ is the surface charge density, $s$  is the distance of the electron avalanche front from the ion disk (assumed frozen during the avalanche development), and $a$  is the disk radius evaluated according to the equation \ref{eq:TransverseRadius}. The formula \ref{eq:SpaceChargeDisk} is derived in detail in the appendix A. 

The resulting effective electric field, which includes the external field minus the space charge field, is then used to calculate \( \alpha_{\mathrm{eff}} \) and proceed to the next step. This iterative procedure continues until the avalanche reaches the anode, where the total generated charge is evaluated considering the contributions of all the steps.

\subsubsection*{Limitations of the approach}

This model assumes that the development of each single avalanche proceeds independently, without accounting for interactions with other avalanches that may be simultaneously generated in the gas. Furthermore, fluctuations in the gain due to statistical effects or microscopic processes are not considered here, but could be included in future developments to improve the model reliability.

\section{Evaluation of effective Townsend coefficient}
\label{alphaModel}

The effective Townsend coefficient \(\alpha_{\mathrm{eff}}\), the Townsend coefficient $\alpha$ and the attachment coefficient $\eta$  were obtained using the MagBoltz program within the Garfield framework \cite{garfieldpp-manual}. MagBoltz is a Monte Carlo solver for electron transport properties in gas mixtures under various electric fields, pressures, and temperatures.

In the simulation, the gas mixture was set up to match the standard RPC gas composition at room temperature (293.15 K) and atmospheric pressure. A range of electric fields was defined  between 10 kV/cm and 100 kV/cm, over which MagBoltz computed the ionization and attachment coefficients. The code then generated a gas table with the ionization coefficients \(\alpha_{\mathrm{eff}}(E)\) corresponding to this range of fields. 

Finally,  the computed \(\alpha_{\mathrm{eff}}\) values were scaled by a factor of 1.43 to normalize the simulated \(\alpha_{\mathrm{eff}}\) at 50 kV/cm with the measured value of approximately 10 mm\(^{-1}\) coming from\cite{chiodini}. It should be noted, however, that some more recent measurements \cite{stoccoMes} provide values more consistent with MagBoltz predictions, and they will be considered in the future to optimize the present simulation code.

\begin{figure}[htbp]
    \centering
    \includegraphics[width=0.65\textwidth]{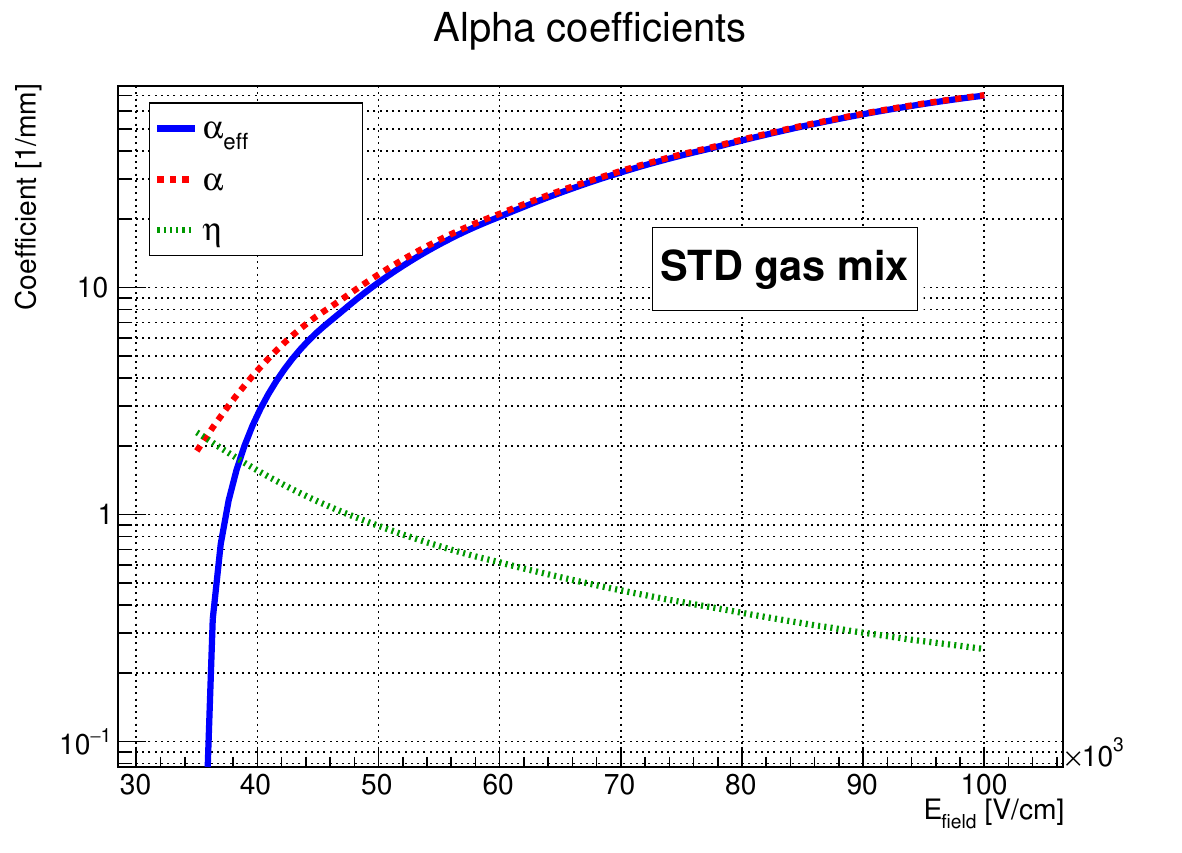}
    \caption{Townsend ($\alpha$), attachment ($\eta$) and effective Townsend ($\alpha_{\mathrm{eff}}$=$\alpha$-$\eta$) coefficient vs Electric field as evaluated by MagBoltz and than rescaled by a factor 1.43 for the standard RPC gas mixture.}
    \label{fig:alphaMagBoltz}
\end{figure}

These tabulated values were subsequently used in the avalanche simulation to determine the exponential growth rate of electrons at each step, depending on the local electric field and are shown as a function of the Electric field in figure~\ref{fig:alphaMagBoltz} .

\section{Results for a Planar RPC with 2 mm Gas Gap at 10 kV}
\label{sec:RPCsim}
To validate the  algorithm, the code has been applied to a planar RPC with a gas gap of 2 mm, filled with the standard RPC gas mixture, and operated at a voltage of 10 kV. The gas gap was discretised into 1000 steps to model the avalanche development starting from the initial position \( x=0 \).

\begin{figure}[htbp]
    \centering
    \includegraphics[width=0.49\textwidth]
    {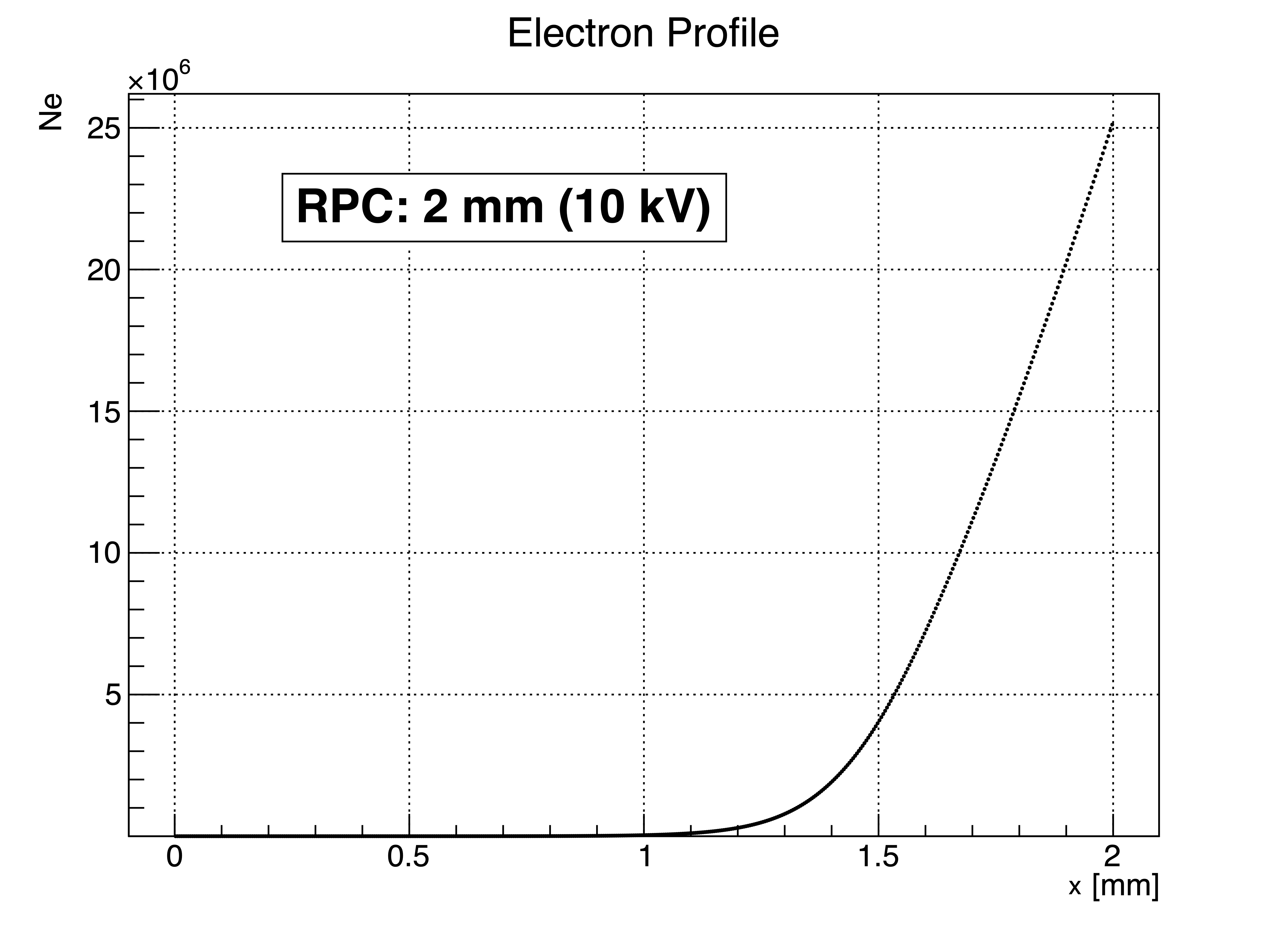}
    \includegraphics[width=0.49\textwidth]{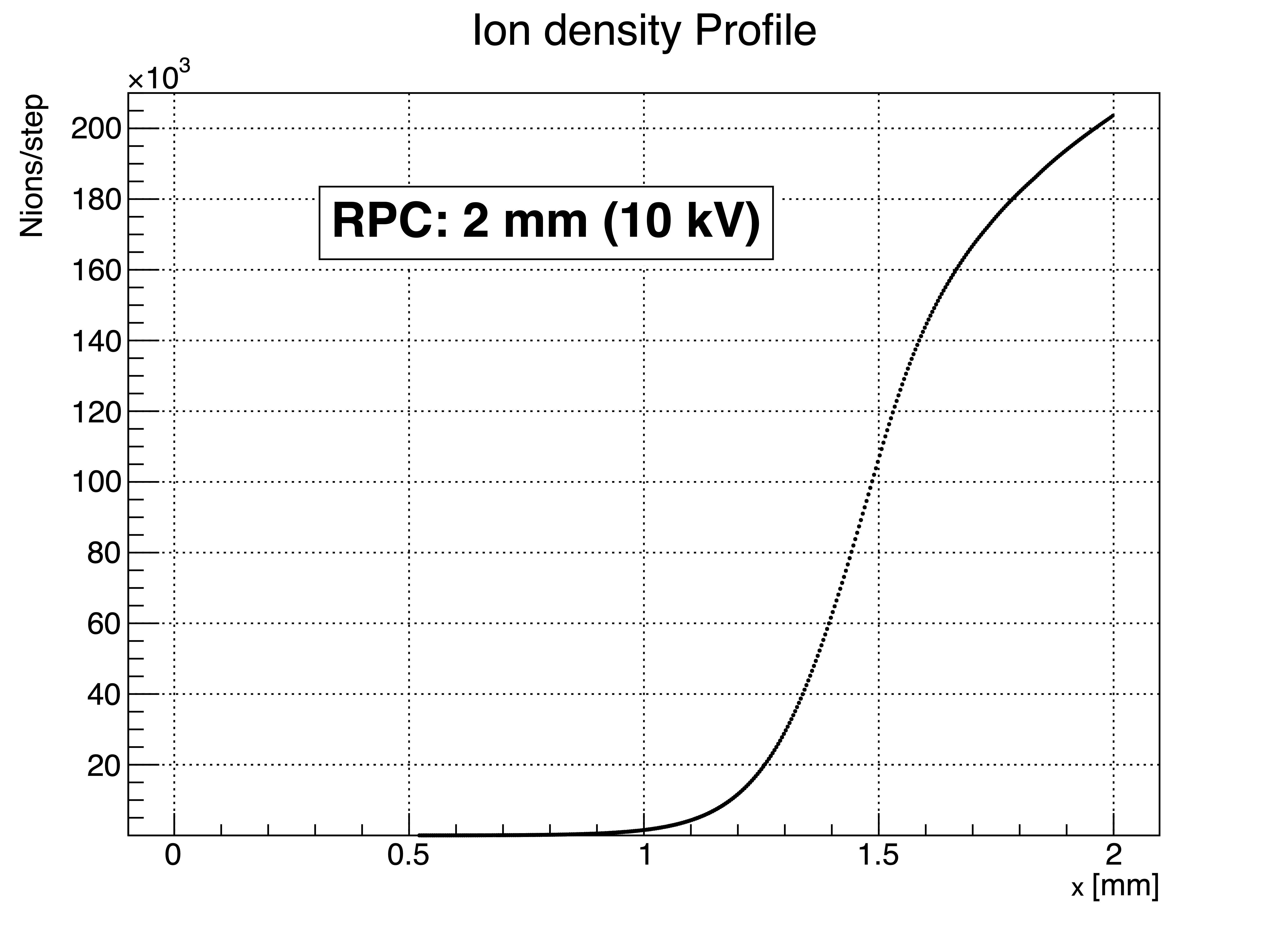}
    \caption{Left plot: Evolution of the number of electrons generated during the avalanche development starting at \( x=0 \). The horizontal axis shows the distance \( x \) in mm covered by the avalanche since the generation, and the vertical axis represents the number of electrons on the avalanche front at each step, illustrating the evolution of the avalanche under the influence of the uniform external electric field and the space charge field. Right plot: Profile of the positive ion density left along the avalanche path upon arrival at the anode. The ions are assumed stationary during the avalanche time scale, thus providing a snapshot of the space charge distribution at the end of the avalanche development.}
    \label{fig:electronIon_evolution}
\end{figure}

Figure~\ref{fig:electronIon_evolution} left, shows the evolution of the electron number in the avalanche as it propagates through the gas gap. The calculation accounts for both the external uniform electric field and the modifying effect of the space charge field generated by the ions left behind. After an exponential growth the electron gain becomes linear due to the space charge contribution.

Figure~\ref{fig:electronIon_evolution} right presents the positive ion profile accumulated along the gap when the avalanche reaches the anode, illustrating the ion distribution responsible for the space charge effects.

\begin{figure}[htbp]
    \centering
    \includegraphics[width=0.65\textwidth]{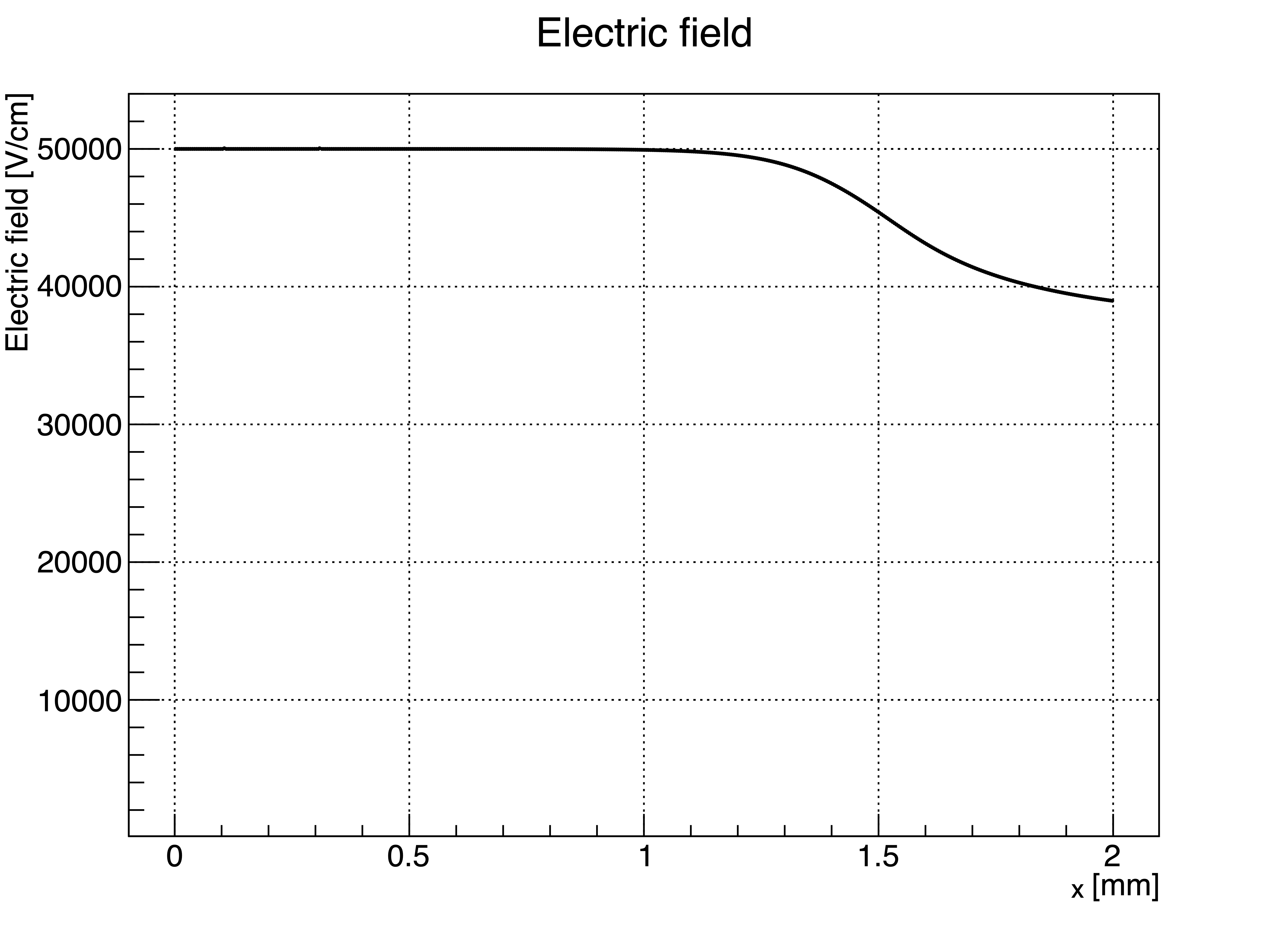}
    \caption{Electric field experienced by the electron cloud as it moves from \( x_0=0 \) to \( x=2 \) mm, considering both the external field and the field due to space charge. The reduction of the effective electric field near the anode due to space charge effects limits the exponential growth of the electron avalanche.}
    \label{fig:electric_field_profile}
\end{figure}

Finally, Figure~\ref{fig:electric_field_profile} illustrates the electric field profile along the gas gap as seen at the front of the electron avalanche, showing the combined effect of the uniform external field and the opposing field generated by the positive ion cloud. The space charge reduces the effective field near the anode, thus limiting the exponential growth phase of the electron avalanche.

\section{Simulation of a Cylindrical RCC Geometry}
\label{RCCsim}

In this section the simulation is applied to a Resistive Cylindrical Chamber (RCC) with electrode thickness of 1 mm, internal radius \( r_i = 2\ \mathrm{mm} \) and external radius \( r_o = 3\ \mathrm{mm} \), filled with the standard RPC gas mixture and operated at 5800 V. 

In the case of a cylindrical Resistive Chamber (RCC), the electric field is not uniform as in the planar geometry but follows the radial dependence dictated by the coaxial electrode configuration. For a potential difference $V$ applied between the inner electrode of radius $r_i$ and the outer electrode of radius $r_o$, the electric field magnitude at a radial position $r$ is given by:
\[
E(r) = \frac{V}{r \, \ln\left( \frac{r_o}{r_i} \right)}.
\]
 As a consequence, depending on the polarity configuration, an electron can move towards regions of either decreasing or increasing electric field, strongly affecting the avalanche development.

The simulation results are initially presented without applying space charge effects, in order to isolate the influence of the purely geometric field gradient on the avalanche development. 

As discussed in section \ref{sec:RPCsim} for the plana racse, three plots are provided:
\begin{enumerate}
    \item The evolution of the number of electrons as a function of the distance \( r \).
    \item The distribution of positive ions left along the avalanche path at the time the electrons reach the anode (ions are assumed immobile during the avalanche timescale).
    \item The electric field profile seen by the avalanche electrons front as they drift from their initial position to the anode.
\end{enumerate}

The study was performed for the two possible polarities:
\begin{itemize}
    \item \textbf{Negative polarity}: The inner electrode is set at negative voltage with respect to the outer one and acts as the cathode. In this configuration, an electron generated near the inner electrode (\( r = 2\ \mathrm{mm} \)) drifts outward toward the anode, moving through a region where the electric field decreases with distance from the center.
    \item \textbf{Positive polarity}: the inner electrode is the anode and the outer electrode is the cathode. Here, an electron generated close to the outer electrode (\( r = 3\ \mathrm{mm} \)) moves inward toward the anode, experiencing an increasing electric field along its trajectory.
\end{itemize}

\begin{figure}[h]
    \centering
    \includegraphics[width=0.48\textwidth]{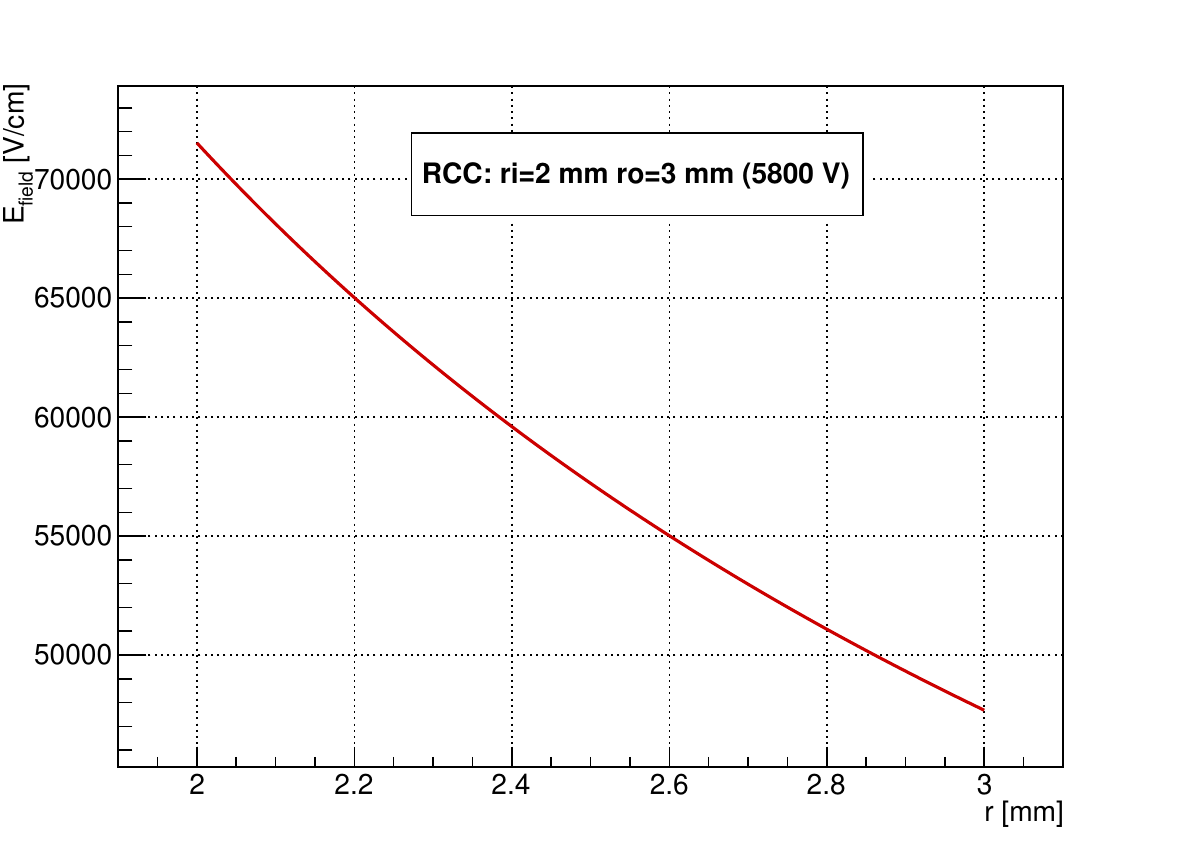}
    \includegraphics[width=0.48\textwidth]{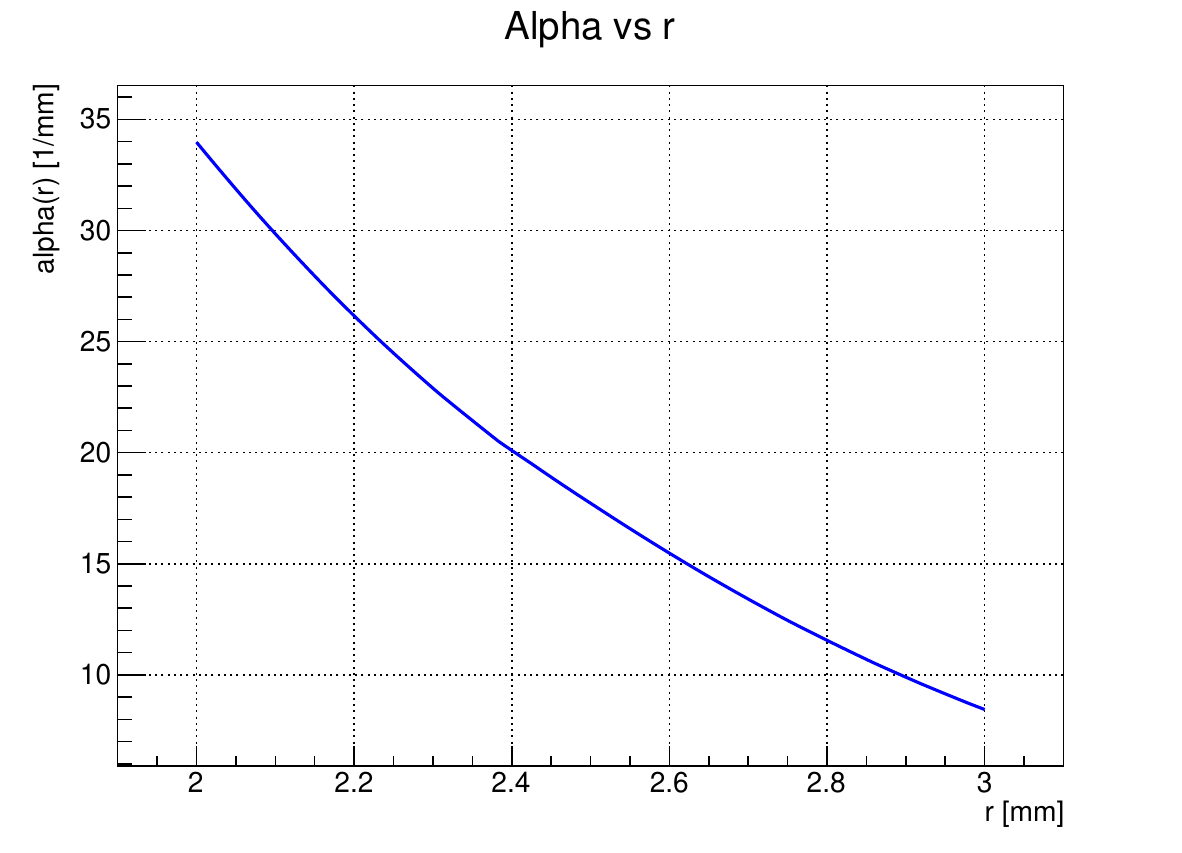}
    \caption{Electric field (left) and effective Townsend coefficient (right) as a function of the radial position for the RCC operated at 5800 V.}
    \label{fig:Efield_cyl_nospace}
\end{figure}

\begin{figure}[h]
    \centering
    \includegraphics[width=0.48\textwidth]{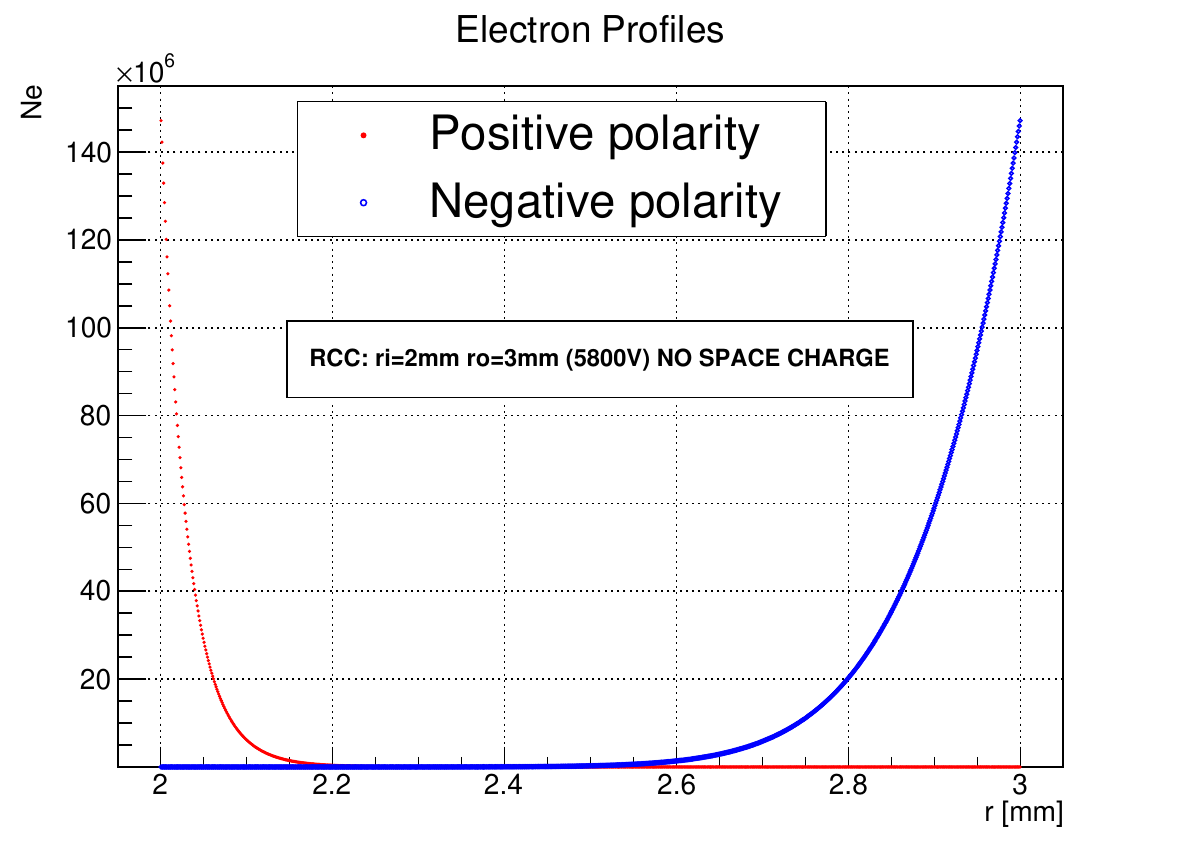}
    \includegraphics[width=0.48\textwidth]{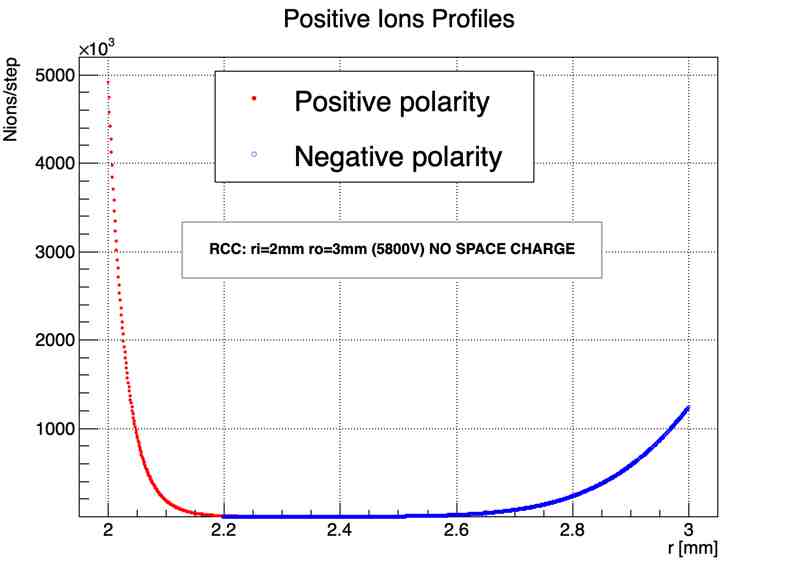}
    \caption{Evolution of the number of electrons along the avalanche path for the cylindrical RCC in both negative and positive polarity (left) and profile of positive ion density left along the avalanche path at the time of arrival at the anode (right) without space-charge effects included.}
    \label{fig:ne_cyl_nospace}
\end{figure}

Figure \ref{fig:Efield_cyl_nospace} shows the value of the electric field (left) and of the Townsend effective parameter (right) $\alpha_{eff}$ as a function of the radial position inside the RCC gas gap, when no space charge is considered and a voltage of 5800 V is applied between the electrodes.
Figure \ref{fig:ne_cyl_nospace} shows the results in terms of electron evolution (left) and ion density profile (right) for both polarities when no space charge is applied.

A second set of plots is generated with the same quantities when space charge effects are included in the simulation. In this case, the electric field is noticeably distorted in the vicinity of the anode due to the accumulation of positive ions, altering the avalanche growth rate as in the planar case. This deformation is more pronounced in regions of higher electric field and is strongly dependent on the chamber polarity.

\begin{figure}[h]
    \centering
    \includegraphics[width=0.48\textwidth]{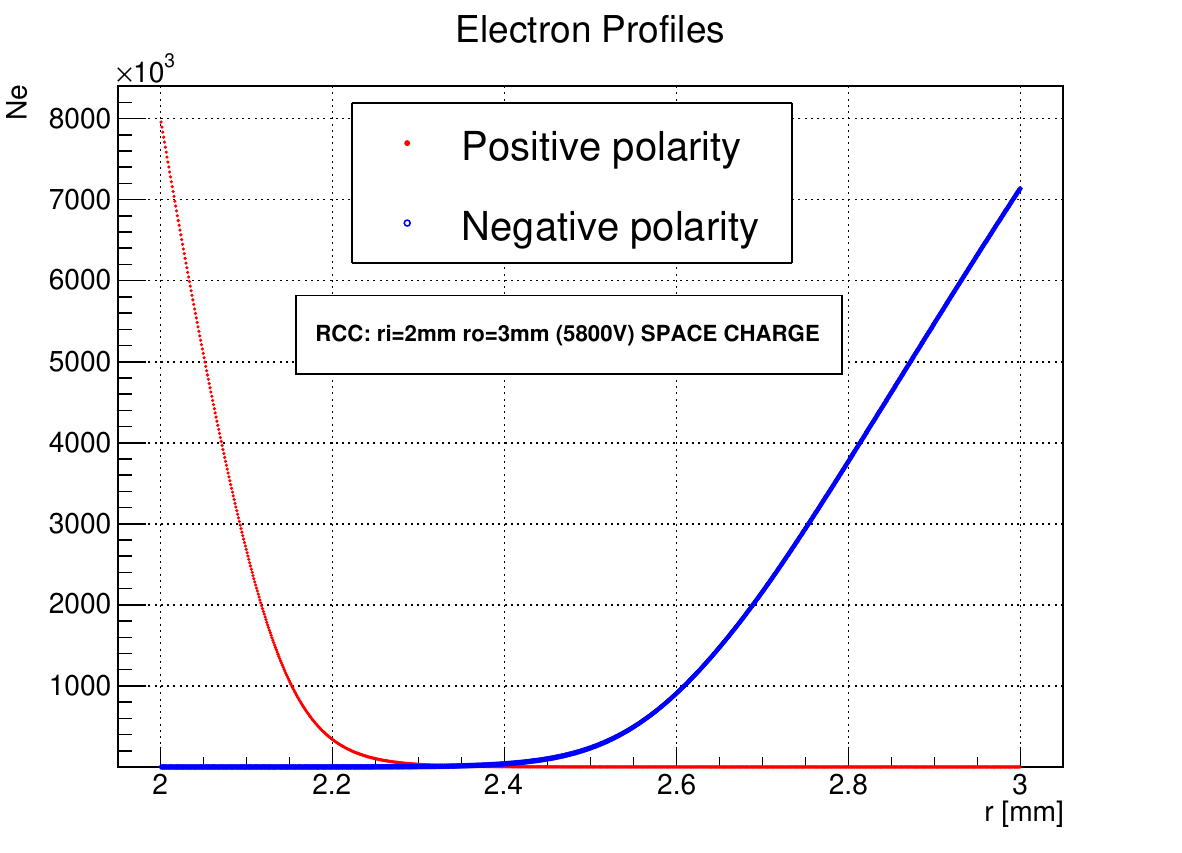}
    \includegraphics[width=0.48\textwidth]{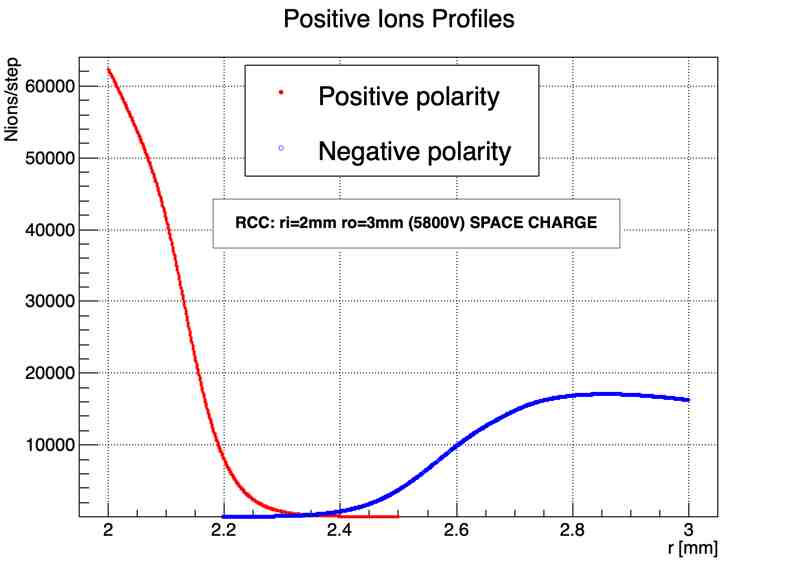}  
    \caption{Evolution of the number of electrons along the avalanche path (left) and ion profile (right) for the cylindrical RCC in both polarities with space-charge effects included.}
    \label{fig:ne_cyl_space}
\end{figure}

\begin{figure}[h]
    \centering
    \includegraphics[width=0.55\textwidth]{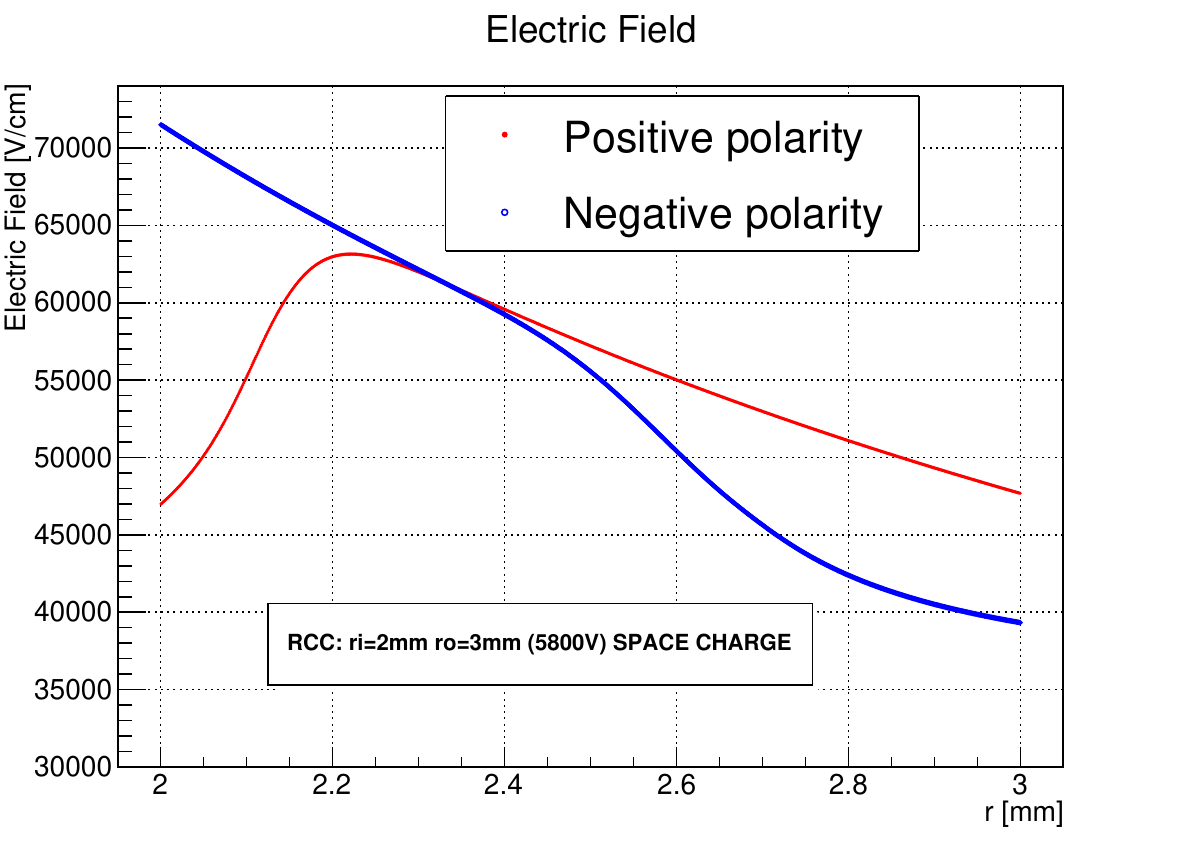}
    \caption{Total electric field (external + space charge) seen by the avalanche front for the cylindrical RCC geometry in the two polarities.}
    \label{fig:Efield_cyl_space}
\end{figure}

Figures \ref{fig:ne_cyl_space} and \ref{fig:Efield_cyl_space} show the results when space charge is applied.

\section{General Framework for Induced Charge Calculation}
\label{inducedCharge}

The calculation of the induced charge on an electrode due to the motion of charges within a detector volume is fundamentally based on the \emph{Ramo-Shockley theorem}. According to this theorem and following the formalization given in \cite{AbbresciaPeskovFonte2018}, the instantaneous current \( i(t) \) induced on an electrode by a charge \( q \) moving with velocity \( \mathbf{v}(t) \) is given by:

\begin{equation}
i(t) = q \, \mathbf{v}(t) \cdot \mathbf{E}_w(\mathbf{r}(t)),
\end{equation}

where \(\mathbf{E}_w(\mathbf{r})\) is the so-called \emph{weighting field}, a hypothetical electric field calculated by setting the electrode of interest at unit potential and all other electrodes grounded, neglecting space charges.

Integrating this current over time yields the total induced charge \( Q_{\rm ind} \):

% Carica indotta (tempo)
\begin{equation}
Q_{\rm ind} = \int i(t)\, dt 
            = q \int n(t)\, \mathbf{v}(t)\!\cdot\! \mathbf{E}_w\!\big(\mathbf{r}(t)\big)\, dt.
\label{ramostokes}
\end{equation}

When the charge moves along a known trajectory, the induced charge can also be expressed as an integral over the spatial coordinate along the path, utilizing the relationship \( \mathbf{v} = \frac{d\mathbf{r}}{dt} \), leading to:

% Cambio di variabile: d\mathbf{r} = \mathbf{v}\,dt
\begin{equation}
Q_{\rm ind} = q \int_{\mathcal{C}} n(\mathbf{r})\, \mathbf{E}_w(\mathbf{r})\!\cdot\! d\mathbf{r}.
\end{equation}

This formulation allows the calculation of induced charge from the knowledge of the charge path and the weighting field, without explicit dependence on time. The weighting field depends solely on the detector geometry and electrode configuration and is independent of the actual electric field and charge distribution within the detector.

In our simulation, we discretize the charge motion into small steps and calculate the incremental induced charge at each step, summing these contributions to obtain the total induced charge. Appendix B explains in detail how the discretization method is implemented in the code.

At each step we weight the contribution with the following weighting field (detailed evaluation of these formulae can be found in appendix B1):
\begin{equation}
E_w = \frac{1}{g + 2 d/\epsilon_r}.
\label{eq:weight_planar}
\end{equation}
for the RPC planar case.

For the cylindrical case the formula is the following:

\begin{equation}
 \; E_w(r) \;=\; \frac{1}{\,r\,\left\{ \displaystyle 
\ln\!\frac{r_o}{r_i} \;+\; \frac{1}{\varepsilon_r}\Big[ 
\ln\!\frac{r_i}{r_i-d} - \ln\!\frac{r_o}{r_o+d}
\Big]\right\} } \; . \; 
\label{eq:Ew_exact_1}
\end{equation}
with inner radius $r_i$, outer  radius $r_o$ and electrode thickness $d$.
 
\begin{figure}[htbp]
    \centering
    \includegraphics[width=0.65\textwidth]{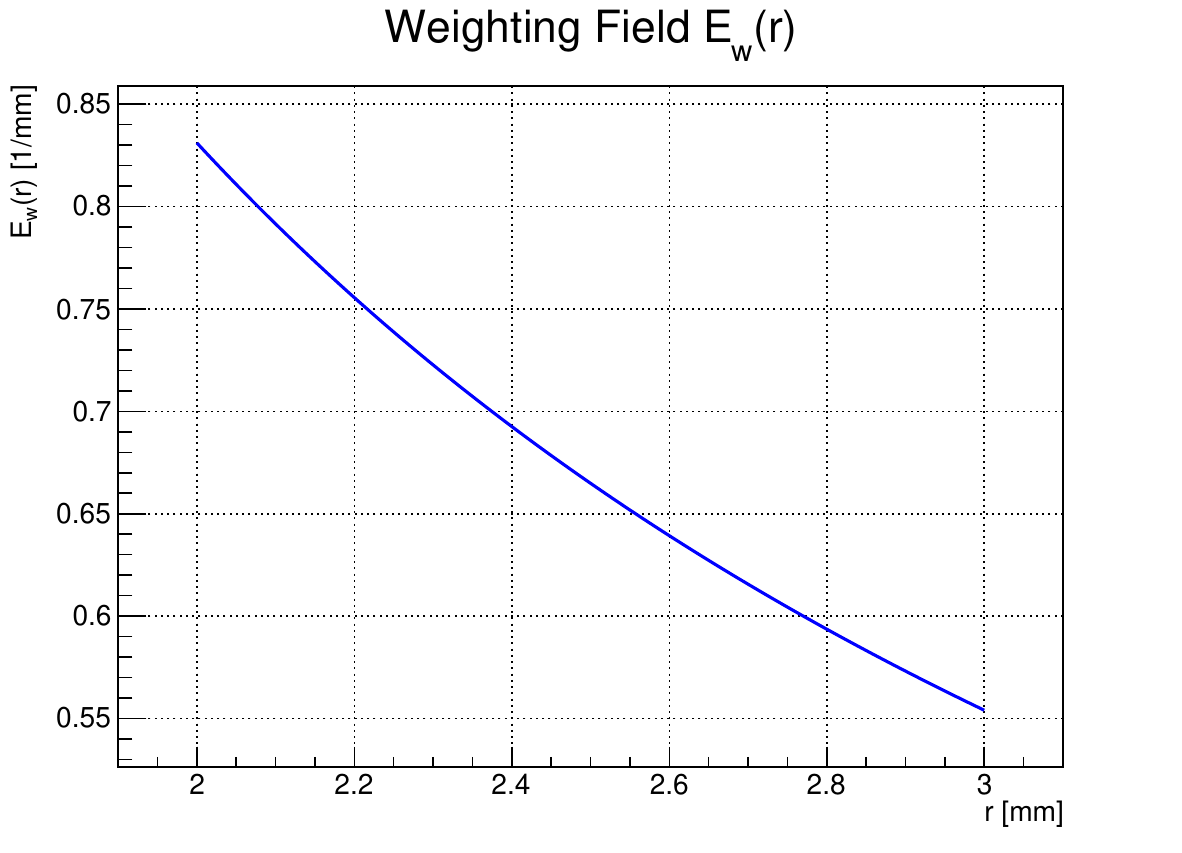}
    \caption{Weighting field $E_w(r)$ for $r_i = 2~\mathrm{mm}$, 
             $r_o = 3~\mathrm{mm}$ and $d = 1~\mathrm{mm}$, computed 
             according to eq.~\ref{eq:Ew_exact_1}. 
             }
    \label{fig:Ew_cyl}
\end{figure}

 As an example, in fig.~\ref{fig:Ew_cyl} we report the profile of the weighting field 
$E_w(r)$ computed according to eq.~\ref{eq:Ew_exact_1} for a cylindrical
geometry with inner radius $r_i = 2~\mathrm{mm}$, outer radius 
$r_o = 3~\mathrm{mm}$ and electrode thickness $d = 1~\mathrm{mm}$.
The curve shows the expected $1/r$-like behavior modulated by the logarithmic
and dielectric terms in eq.~\ref{eq:Ew_exact_1}, resulting in a weighting
field that spans from values lower to higher with respect to the weighting field of a planar configuration with
similar gap and electrode size.
For comparison, the weighting field of a planar RPC with gas gap and electrode thickness of 1 mm is 0.71 (for $\epsilon_r$=5).

\subsection{Induced Charge in Realistic Detector Configurations}
\label{InducedCharge_realisticConfigurations}
In this section we explore the induced charge generated in few realistic detector configurations:  

\begin{itemize}
    \item \textbf{RPC} (Resistive Plate Chamber) with electrode thickness 
    $1.8~\mathrm{mm}$ and gas gap of:
    \begin{itemize}
        \item $2~\mathrm{mm}$, with an applied voltage 
    of $10{,}000~\mathrm{V}$ across the gap.
         \item $1~\mathrm{mm}$, with an applied voltage 
    of $6{,}000~\mathrm{V}$ across the gap.
         \item $0.5~\mathrm{mm}$, with an applied voltage of $4{,}200~\mathrm{V}$ across the gap.
    \end{itemize}
    \item \textbf{RCC} (Resistive Cylindrical Chamber) with electrode thickness 
    $1~\mathrm{mm}$, inner radius $r_i = 2~\mathrm{mm}$, outer radius 
    $r_o = 3~\mathrm{mm}$, and operating voltage of $5{,}800~\mathrm{V}$.
\end{itemize}

For the RPC case, we consider the situation where a single electron 
initiates the avalanche at a given distance from the cathode surface.  
The induced charge is calculated following the procedure described in 
Sec.~\ref{inducedCharge}.  

\begin{figure}[ht]
    \centering
    \includegraphics[width=0.48\textwidth]{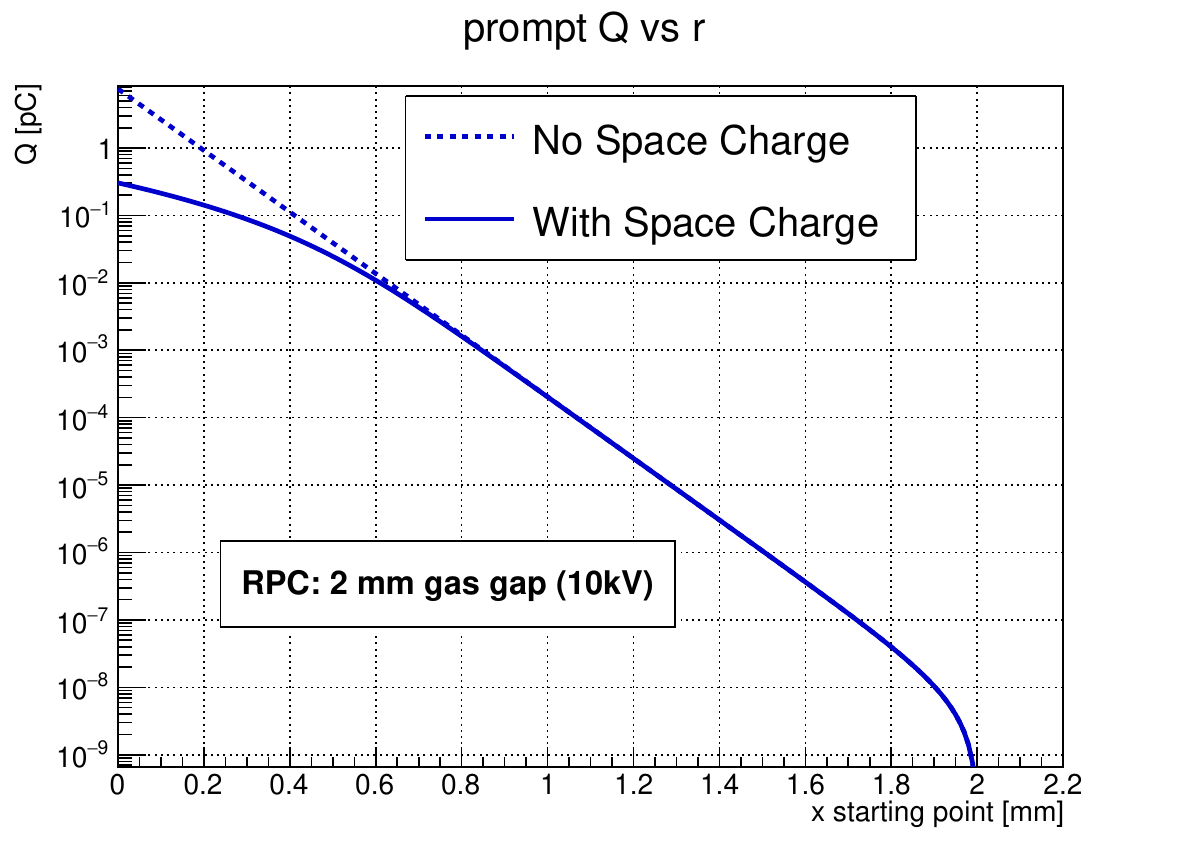}   
    \includegraphics[width=0.48\textwidth]{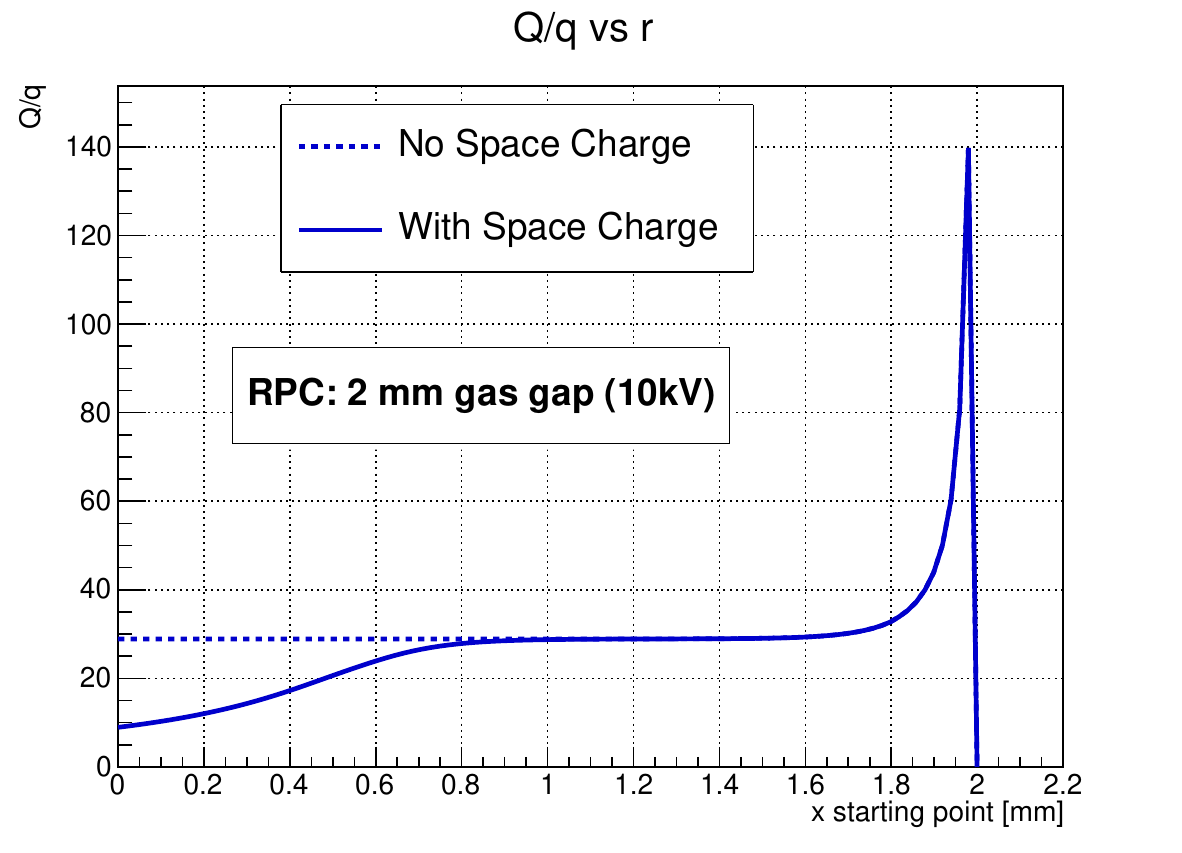}   
    \caption{Induced charge  $Q_{\rm ind}$ and $Q_{\rm tot}/q_{\rm ind}$ for a 2 mm gas gap at 10 kV, with and without space charge.}
    \label{fig:charge2mm}
\end{figure}

\begin{figure}[ht]
    \centering
    \includegraphics[width=0.48\textwidth]{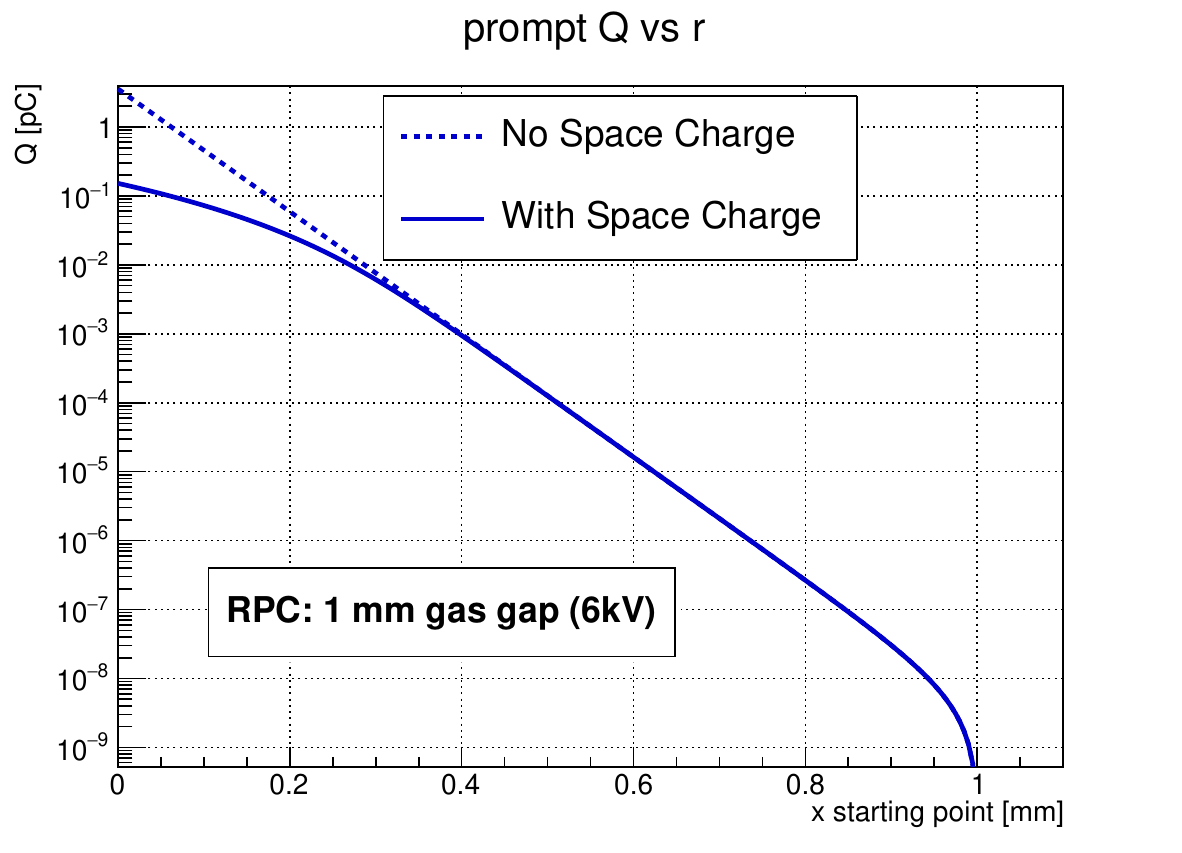}   
    \includegraphics[width=0.48\textwidth]{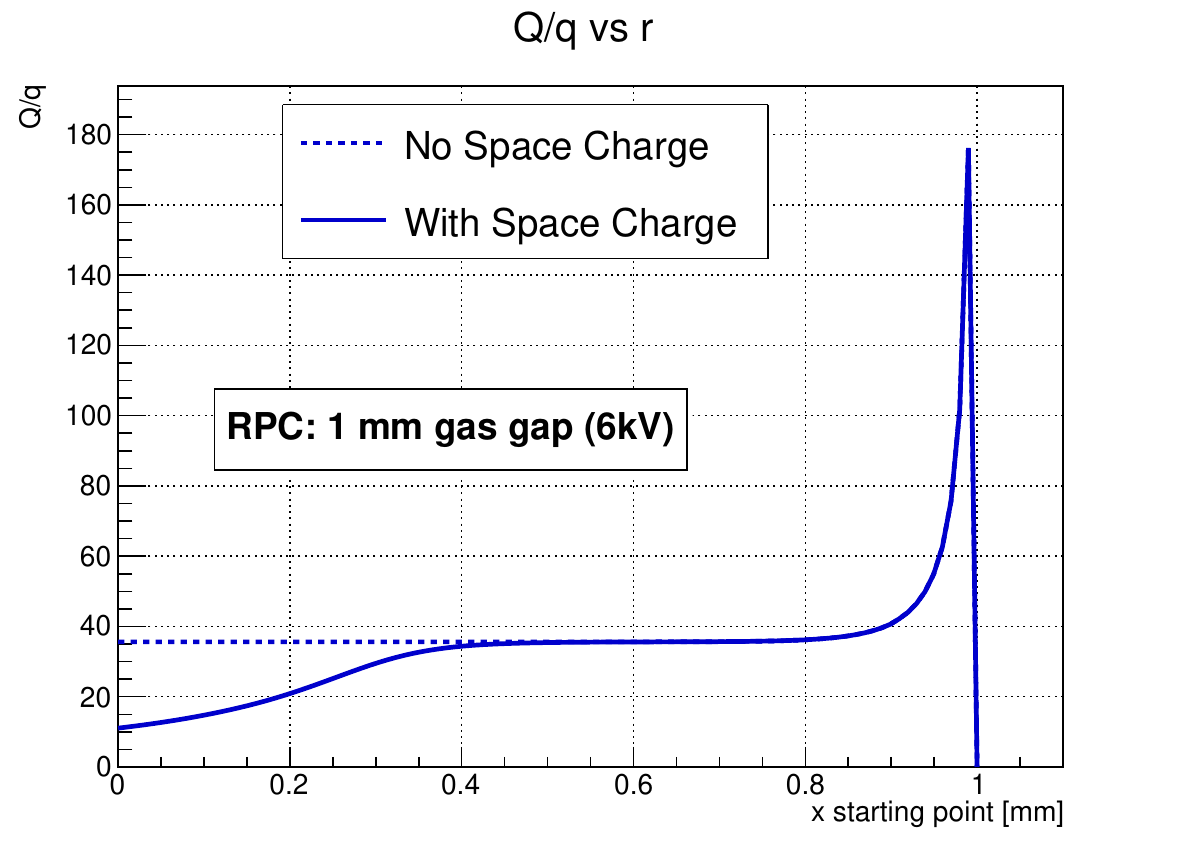}   
    \caption{Induced charge $Q_{\rm ind}$ and $Q_{\rm tot}/q_{\rm ind}$ for a 1 mm gas gap at 6 kV, with and without space charge.}
    \label{fig:charge1mm}
\end{figure}

\begin{figure}[ht]
    \centering
    \includegraphics[width=0.48\textwidth]{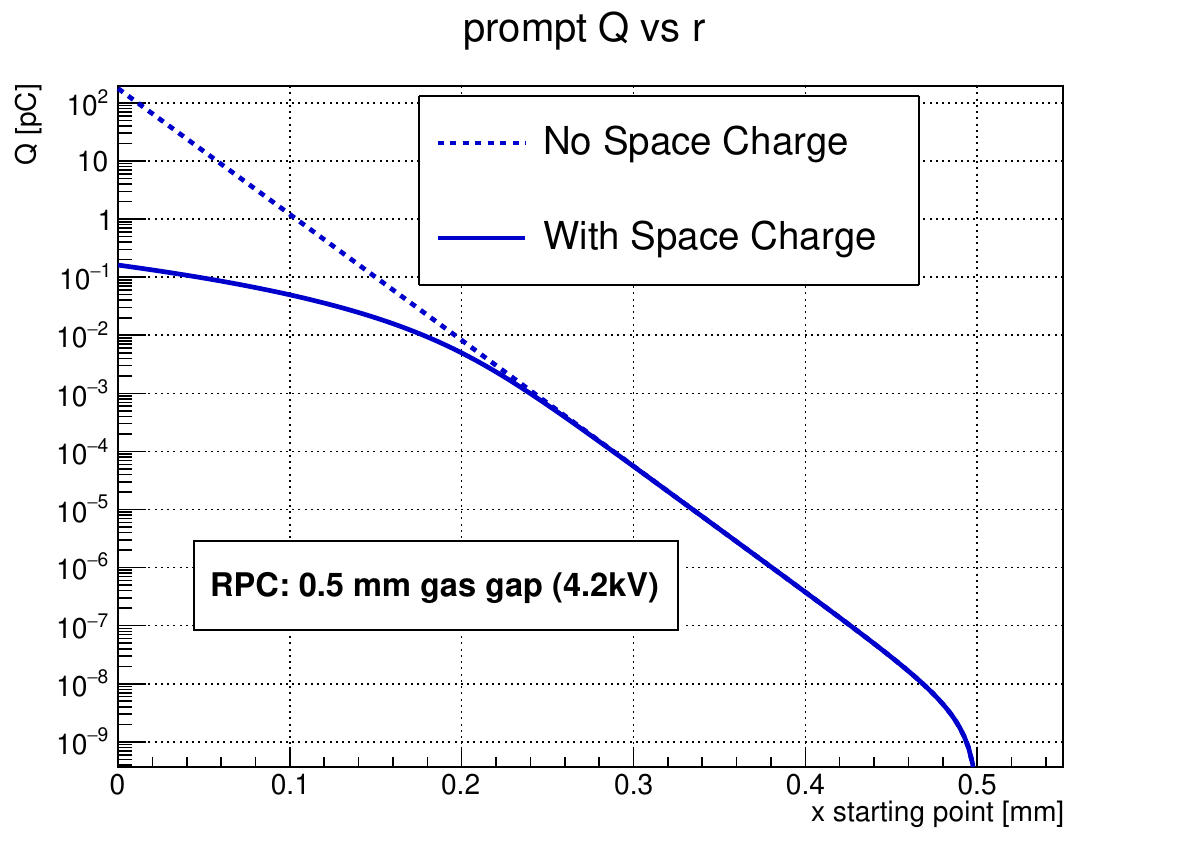}   
    \includegraphics[width=0.48\textwidth]{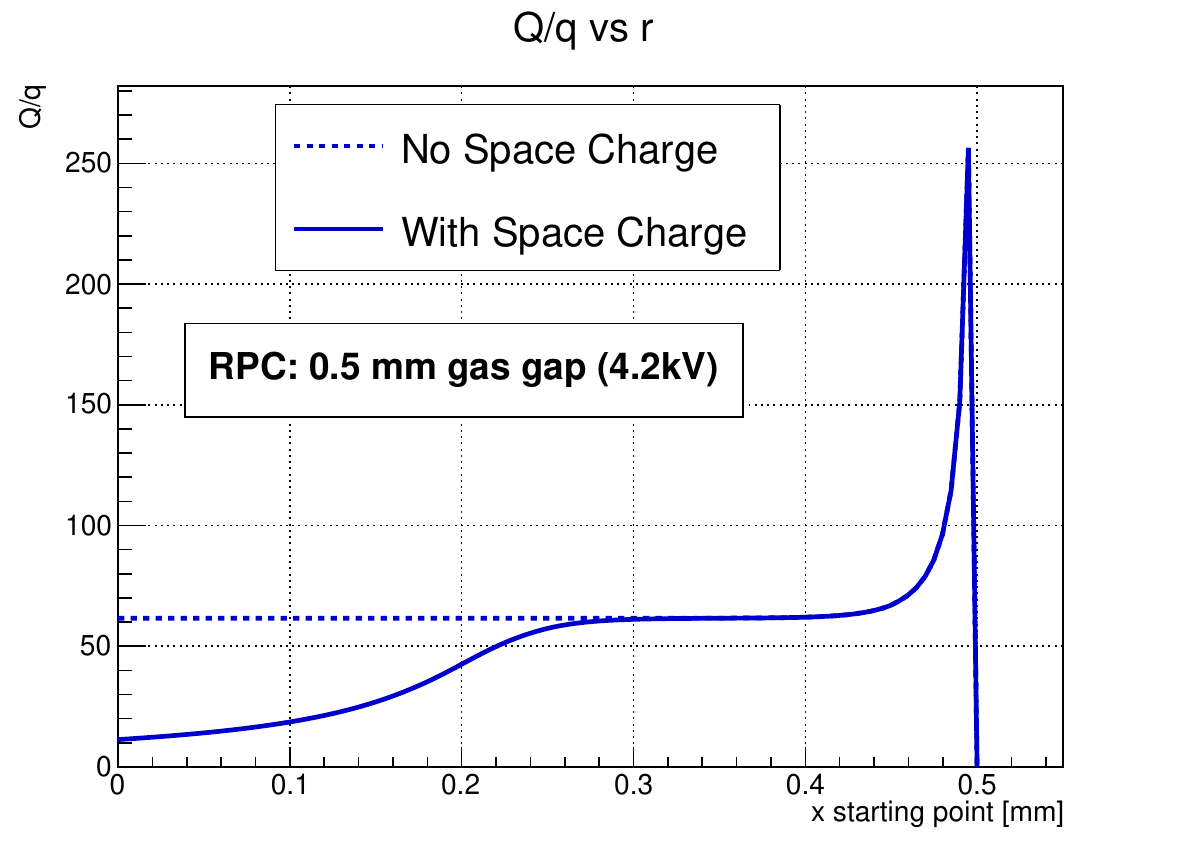}   
    \caption{Induced charge $Q_{\rm ind}$ and $Q_{\rm tot}/q_{\rm ind}$ for a 0.5 mm gas gap at 4.2 kV, with and without space charge.}
    \label{fig:charge05mm}
\end{figure}

In Figs.~\ref{fig:charge2mm}, \ref{fig:charge1mm}, and \ref{fig:charge05mm} the results of the simulation for the three RPC configurations, are presented.
 
For each case, the induced charge (left plot) is shown as a function of the avalanche starting point along the gas gap, both with and without the inclusion of space charge effects. 
In addition, we show the ratio between the total generated charge and the induced charge (right plot).

The comparison across the three gaps clearly shows that the contribution of space charge effects increases as the gas gap decreases according to different RPC geometries. 

In the \emph{absence} of space charge, and for roughly the first \(\sim 90\%\) of the gap (i.e.\ for starts not too close to the anode), the ratio behaves as expected from the Shockley--Ramo picture,
which, for a planar gap with nearly uniform weighting field \(\mathbf{E}_w\) and effective Townsend coefficient \(\alpha_{\rm eff}\), yields the well-known proportionality
\begin{equation}
Q_{\rm ind} \;\propto\; \frac{E_w}{\alpha_{\rm eff}}\!\left(e^{\alpha_{\rm eff} x}-1\right),
\qquad
Q_{\rm tot} \;\propto\; e^{\alpha_{\rm eff} x}
\label{eq:qinduced}
\end{equation}

so that in this regime and for paths large enough to neglect the term $1$ with respect to $ e^{\alpha_{\rm eff} x}$ in the $Q_{\rm ind}$ formula
\[
\boxed{\;\dfrac{Q_{\rm tot}}{Q_{\rm ind}} \;\approx\; \dfrac{\alpha_{\rm eff}}{E_w}\;=\;\text{const.}\;}
\]
Here \(x\) is the available drift/multiplication distance from the starting point to the anode. This explains the approximately flat behavior of \(Q_{\rm tot}/Q_{\rm ind}\) over most of the gap when space charge is neglected.

When space charge is considered, for starting positions \(r\) \emph{near the cathode}, the local space charge density grows in the final part of the drift (close to the anode), so the amplification is partially saturated in that last segment. This saturation suppresses further multiplication but \emph{enhances the induced signal contribution} from carriers already present, because a sizable portion of the avalanche still drifts through the region. As a result the \(Q_{\rm tot}/Q_{\rm ind}\) is reduced.

For starting positions \emph{very close to the anode}, the finite path length becomes small and the term 1 in \(\big(e^{\alpha_{\rm eff} x}-1\big)\) arising from eq.~\eqref{eq:qinduced} becomes important, breaking the simple proportionality above. In practice, as \(x\to 0\) the induced charge no longer scales linearly with the total generated charge, and the ratio \(Q_{\rm tot}/Q_{\rm ind}\) departs from the constant \(\alpha_{\rm eff}/E_w\).

In the case of the RCC, the detector can be operated in two different configurations, depending on whether the cathode is chosen to be the inner cylindrical electrode or the outer one. 

\begin{figure}[htbp]
    \centering
    \includegraphics[width=0.48\textwidth]{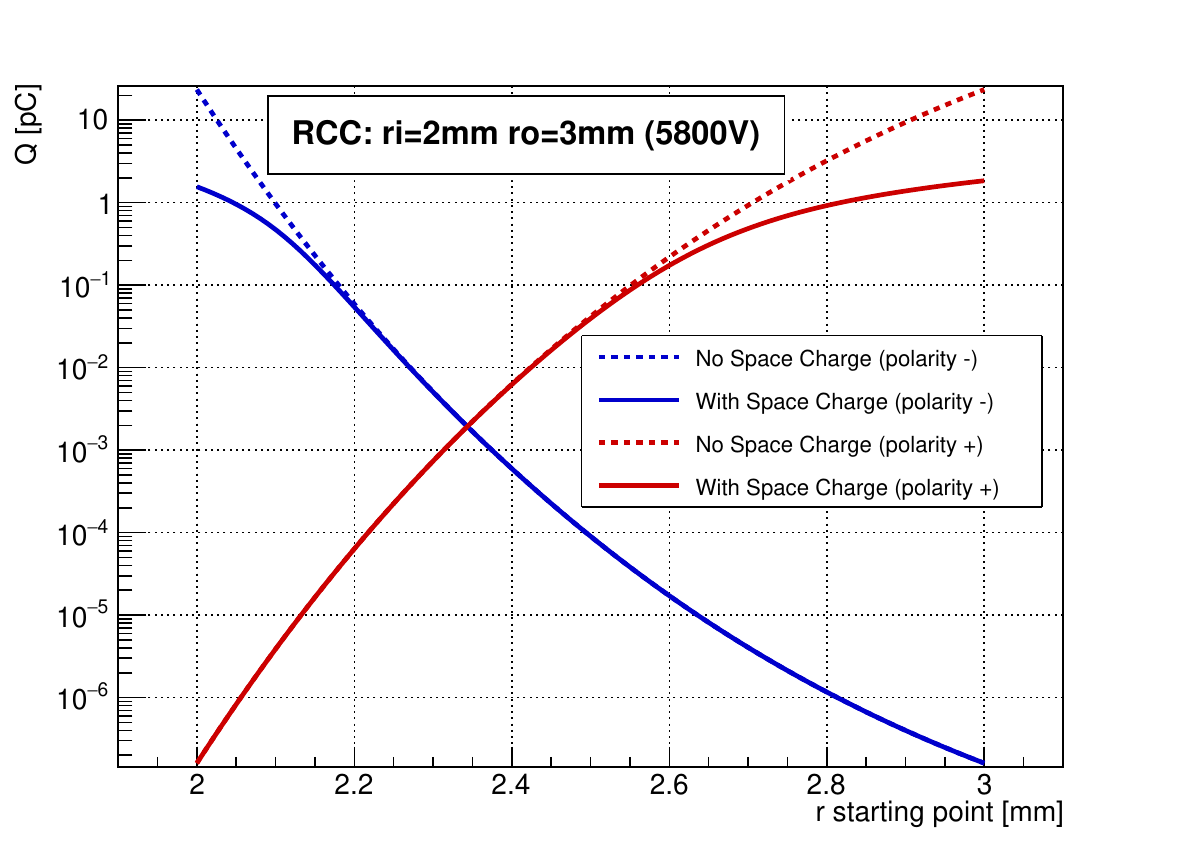}
    \includegraphics[width=0.48\textwidth]{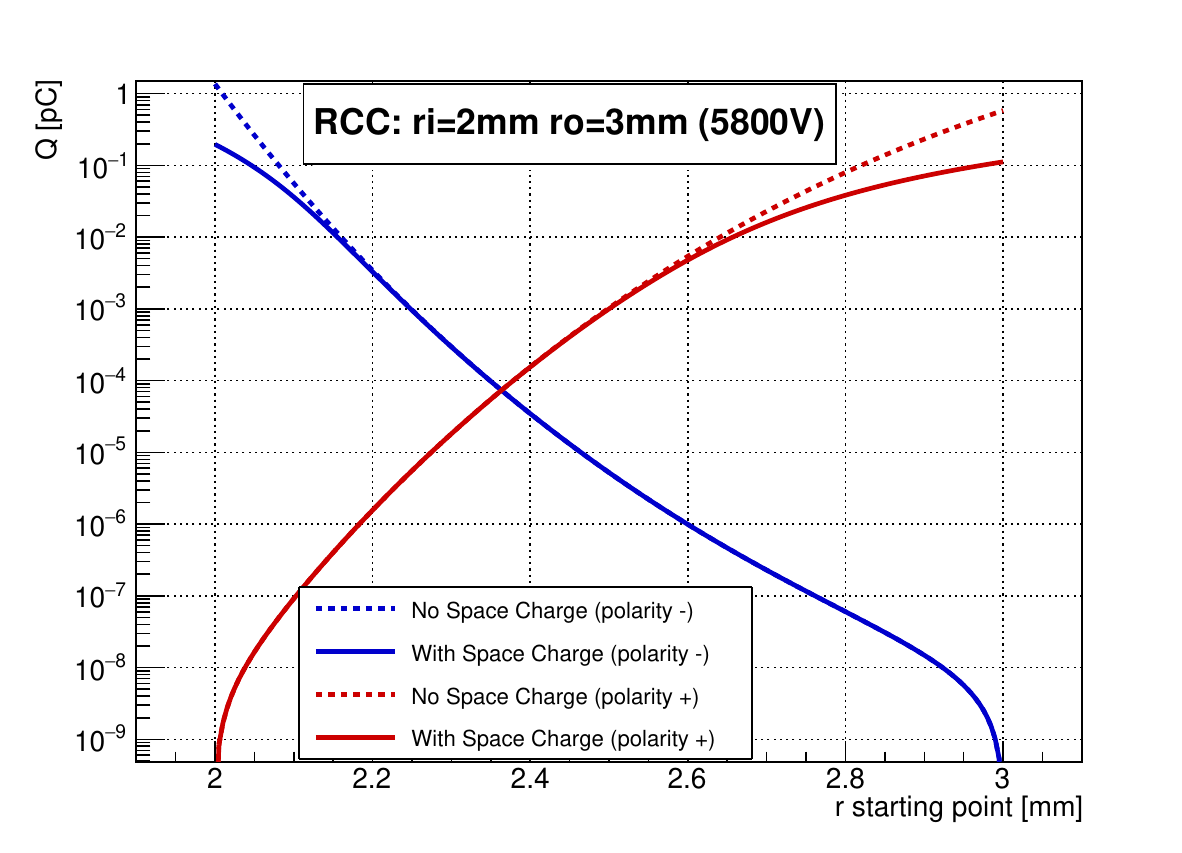}
    % Placeholder for RCC charge plots
    \caption{Total (left) and induced (right) charge for the RCC configuration, as a function of the avalanche starting point, with and without space charge effects, for both polarities and for an applied voltage of 5800 V.}
    \label{fig:RCC_charges}
\end{figure}

Figure~\ref{fig:RCC_charges} shows the total charge (left plot) and the induced charge (right plot) as a function of the avalanche starting position, with and without space charge applied for the two different polarities, and for an operating  voltage of 5800 V: for negative polarity the cathode is the inner electrode, for positive polarity the cathode is the outer.  

The asymmetry between the two polarities is evident.  
In particular, for an avalanche starting at $r = r_i = 2$~mm in negative polarity, the total charge generated upon reaching the anode ($r_o = 3$~mm) is about 1.6~pC, while the induced charge is approximately 0.2~pC.  
In positive polarity, an avalanche starting at $r = r_o = 3$~mm and propagating towards the anode ($r_i = 2$~mm) produces a total charge of about 1.9~pC and an induced charge of roughly 0.12~pC.  

The plots in Figures~\ref{fig:RCC_charges} highlight a clear polarity dependence for the RCC geometry.  
In the absence of space charge, the total charge is the same for both polarities, as expected from the symmetry of the formula for the gain applied to avalanches starting exactly at the cathode position.  
When space charge is included, however, the gain is more strongly reduced in negative polarity.

\begin{figure}[h!]
    \centering
    \includegraphics[width=0.65\textwidth]{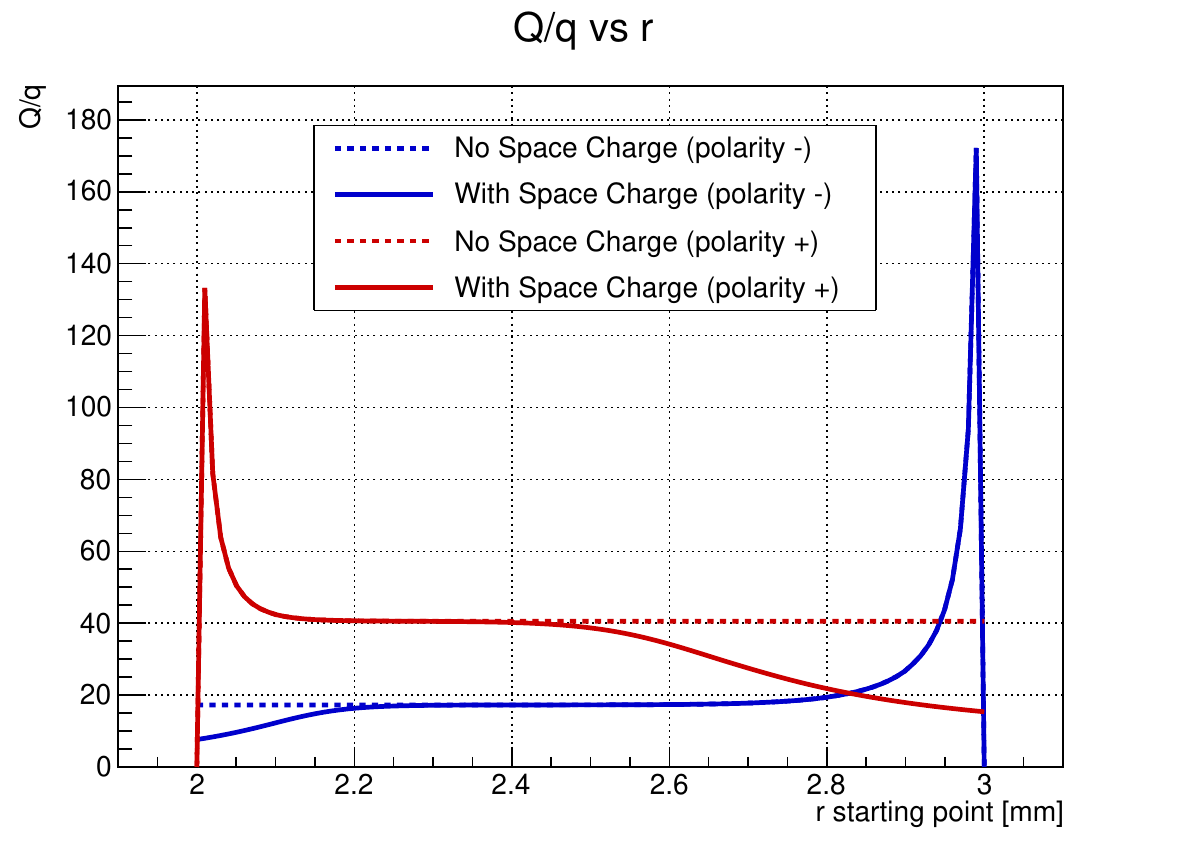}
    \caption{Charge ratio $\frac{Q_{tot}}{q_{induced}}$ vs avalanche starting point, for a RCC detector with $r_i$=2 mm and $r_o$=3 mm, with and without space charge applied and operated at 5.8 kV in both positive and negative polarities.}
    \label{fig:ChargeRatioVsDistanceRCC_2_3}
\end{figure}

Regarding the induced charge, fig.\ref{fig:ChargeRatioVsDistanceRCC_2_3} shows the ratio $\frac{Q_{tot}}{q_{induced}}$ as a function of avalanche starting point, indicating that  the fraction of total to induced charge is larger in positive polarity compared to negative polarity. 

At first sight this may seem counterintuitive, since in positive polarity the weighting field \(E_w(r)\) is larger near the anode, where the avalanche multiplication is stronger.  
However, in positive polarity the avalanche starts in a low-field region and experiences weak multiplication for a large fraction of its path;  
only in the final small region near \(r_i\), where both \(E\) and \(E_w\) are high, does the avalanche grows significantly.  
This short high-gain region contributes to the induced charge for a relatively small fraction of the avalanche development.

In negative polarity, by contrast, the avalanche starts in a high-field region and grows rapidly from the very beginning.  
As a result, a large number of electrons is present for a longer distance (and hence for a longer time) during its drift.  
The longer duration of high-charge transport compensates for the lower \(E_w\) value in the region where higher is the number of electrons, leading to a higher integrated induced charge.

This interplay between avalanche growth dynamics and the spatial profile of the weighting field explains the observed polarity asymmetry.  

One more consideration has to be done however. If all avalanches would start at the farthest point from the anode, one would generally expect positive polarity to produce a larger total charge but a smaller induced charge with respect to negative polarity.  
In realistic cases, however, the primary ionizations are distributed along the gap, and as seen in Figure~\ref{fig:RCC_charges} right,  
in negative polarity only primary electrons generated very close to the cathode produce a significant induced charge,  
while in positive polarity the range of starting positions yielding a reasonable induced charge is broader.  
This suggests a potentially higher detection efficiency in positive polarity when realistic primary-ionization distributions are taken into account.

\section{Primary ionization distributions in the standard gas mixture}
\label{ionizationModel}

In order to evaluate the realistic distributions of primary ionization clusters and the number of secondary electrons per cluster for the standard RPC gas mixture, we used the programs \textsc{Heed} and \textsc{Magboltz} within the \textsc{Garfield++} framework.  
The simulation code initializes the gas properties and geometry, and generates 
muon tracks crossing the gap along the \(z\)-direction.  
The \texttt{TrackHeed} class is then used to simulate the ionization clusters along the muon path.  
For each event, the code counts the number of primary clusters per unit length and records the multiplicity of electrons in each cluster.  

\begin{figure}[h!]
    \centering
    \includegraphics[width=0.48\textwidth]{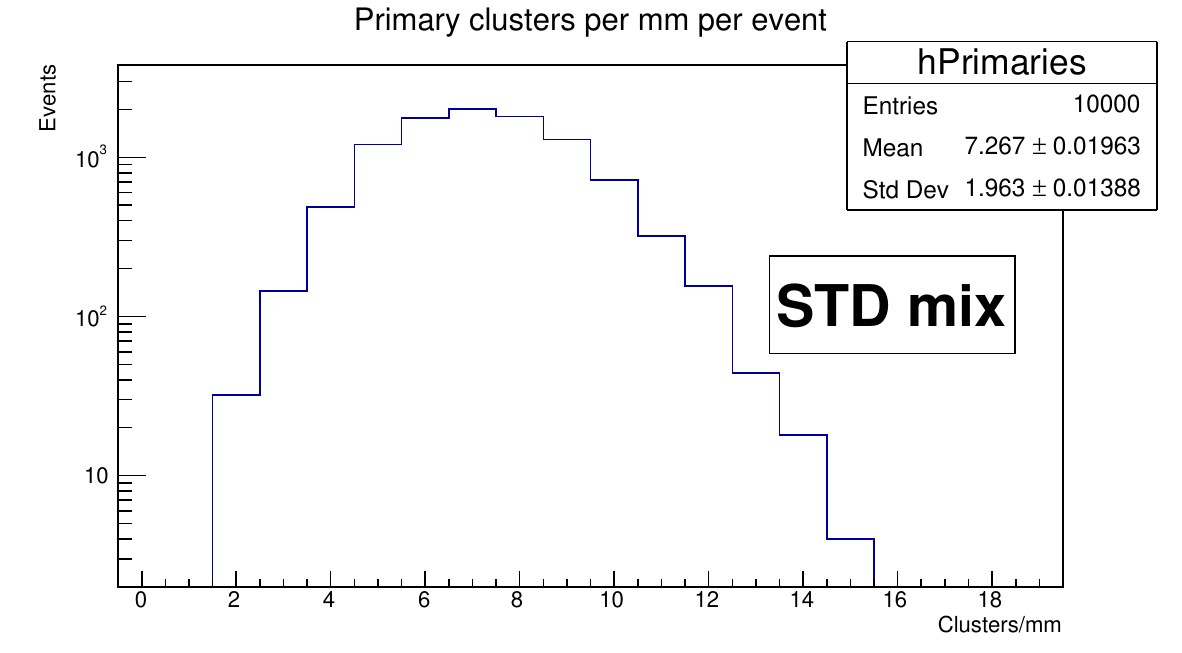}
\includegraphics[width=0.48\textwidth]{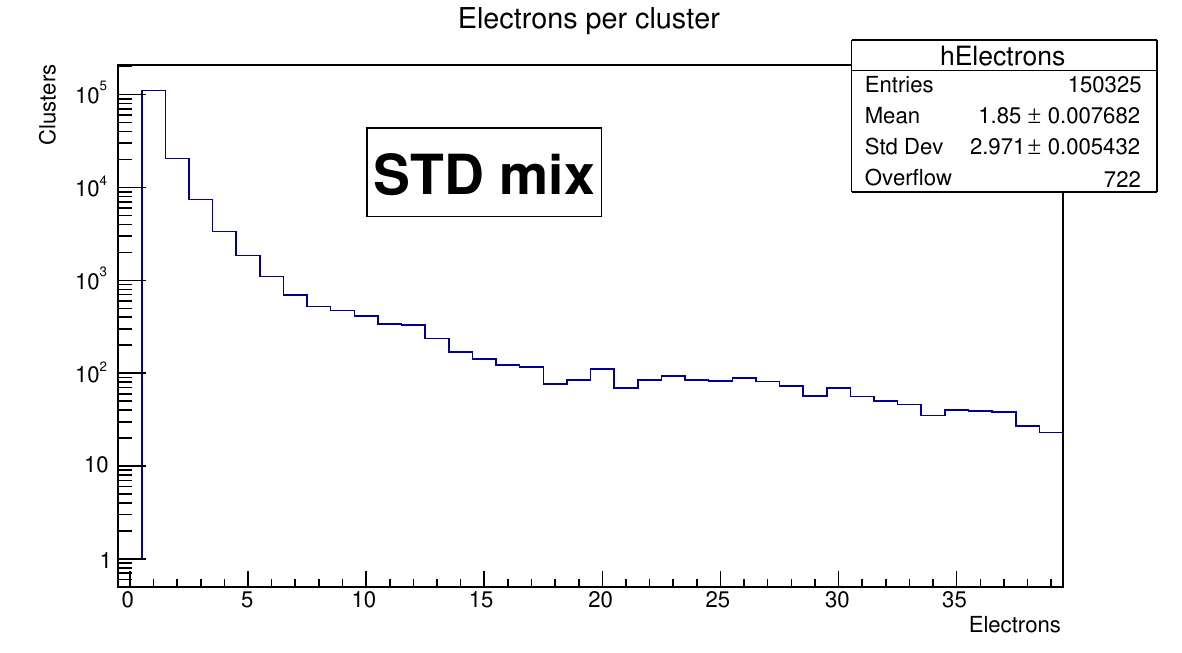}
    \caption{Ionization distributions obtained with \textsc{Heed} for the standard RPC gas mixture simulated on a gas gap of 2 mm. 
    Left: number of primary clusters per mm. Right: number of electrons per primary cluster.}
    \label{fig:heed_clusters}
\end{figure}

Figure~\ref{fig:heed_clusters} shows the two distributions for the standard gas mixture and simulated on a RPC with gas gap of 2 mm.  
The left panel corresponds to the distribution of the number of primary clusters per mm, while the right panel shows the distribution of the number of electrons per cluster.

From the resulting histograms we extract:
\begin{itemize}
    \item The mean number of primary clusters produced per millimetre: 
    \(\langle N_{\rm clusters} \rangle \approx 7.3\ \mathrm{mm^{-1}}\).
    \item The probability distribution of the number of electrons per primary cluster, 
    with a mean value \(\langle N_{e^{-}} \rangle \approx 1.85\).
\end{itemize}

\section{Full Avalanche Simulation with Realistic Primary Ionization}
\label{FullSim}

In this section we assemble all ingredients introduced so far into a full event–by–event simulation and apply it to selected planar RPC configurations with different gas gaps and to RCC geometries with cylindrical symmetry. At this stage, simulated tracks traverse the gas gap perpendicularly in the planar RPC case and radially starting from the cylinder center in the RCC case.

For each event, we first generate the number of primary ionization clusters from the distribution shown in Fig.~\ref{fig:heed_clusters} (left), rescaled to the actual gas-gap thickness. For every generated cluster, the number of secondary electrons is sampled from the normalized probability distribution of Fig.~\ref{fig:heed_clusters} (right). The primary clusters are then placed at uniformly random positions along the gas gap thickness. From each cluster position, its secondary electrons are launched and each electron independently develops an avalanche according to the stepwise procedure detailed in the previous sections (including space-charge). 

The final event observables are obtained by summing the contributions of all electrons from all clusters. Denoting by \(N_c\) the number of primary clusters in the event, and by \(n_e^{(c)}\) the number of secondary electrons in cluster \(c\), the total generated charge \(Q_{\rm tot}\) and the induced charge \(Q_{\rm ind}\) of the event are

\begin{equation}
Q_{\rm tot}^{\mathrm{(event)}} \;=\; \sum_{c=1}^{N_c} \;\sum_{k=1}^{n_e^{(c)}} \; Q_{\rm tot}^{(c,k)} \,,
\qquad
Q_{\rm ind}^{\mathrm{(event)}} \;=\; \sum_{c=1}^{N_c} \;\sum_{k=1}^{n_e^{(c)}} \; Q_{\rm ind}^{(c,k)} \,,
\label{eq:event_sums}
\end{equation}

where \(Q_{\rm tot}^{(c,k)}\) and \(Q_{\rm ind}^{(c,k)}\) are, respectively, the total and induced charges generated by the \(k\)-th electron of the \(c\)-th cluster, starting from its sampled position and evolving in the local ( space–charge–modified) electric field. 
Equation~\eqref{eq:event_sums} makes explicit that in this simulation the detector response for a single traversing particle is the incoherent sum of many independent microscopic avalanches seeded along the track.

\section{Application of the full Avalanche Simulation to the RPC case}
\label{RPCFullSim}

\begin{figure}[h!]
    \centering
    \includegraphics[width=0.48\textwidth]{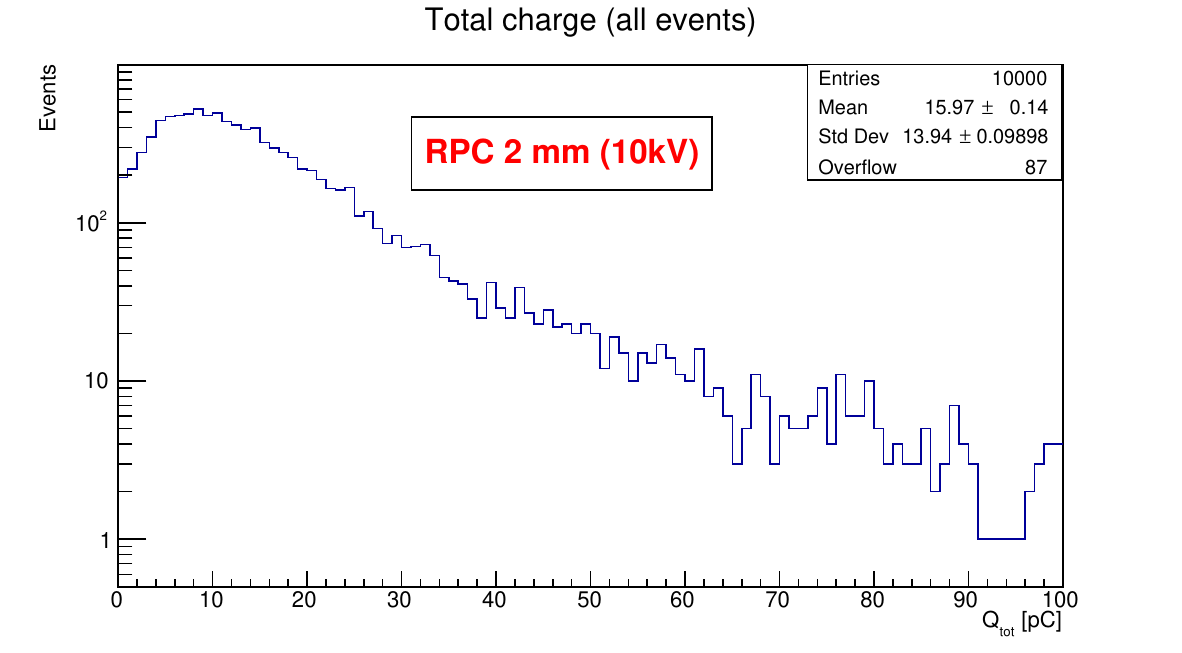}
\includegraphics[width=0.48\textwidth]{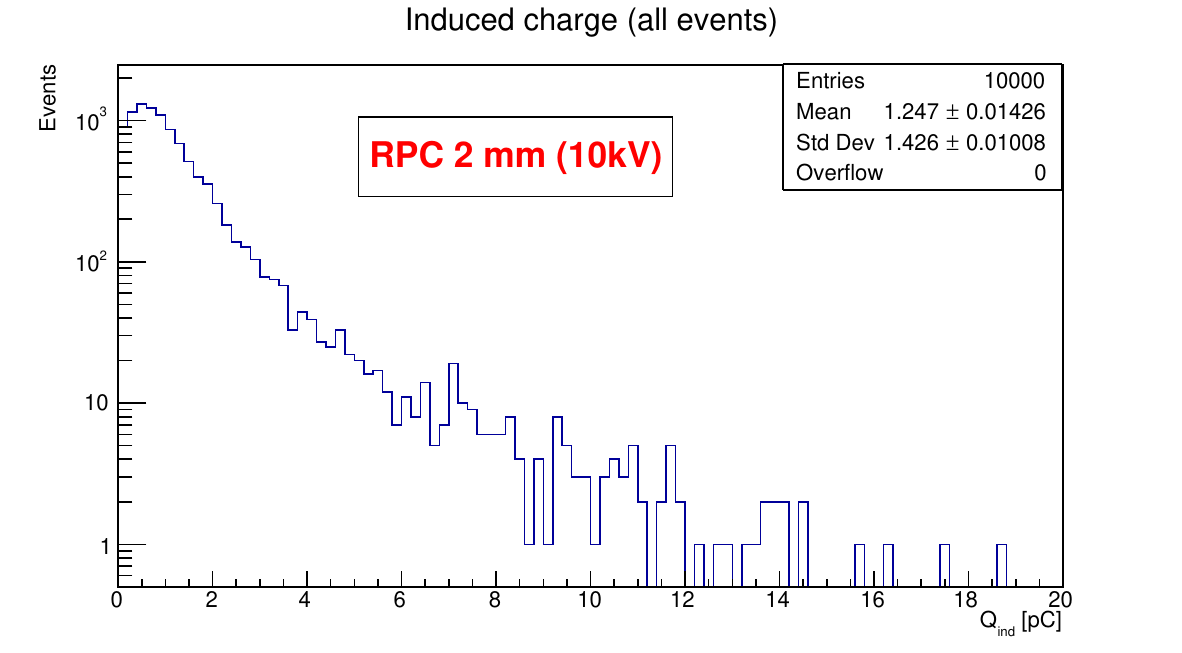}
    \caption{Total charge (left) and induced charge (right) distribution for 2 mm gas gap RPC operated with standard gas mixture at 10kV}
    \label{fig:chargeDistributionsRPC2mm}
\end{figure}

\begin{figure}[h!]
    \centering
    \includegraphics[width=0.49\textwidth]{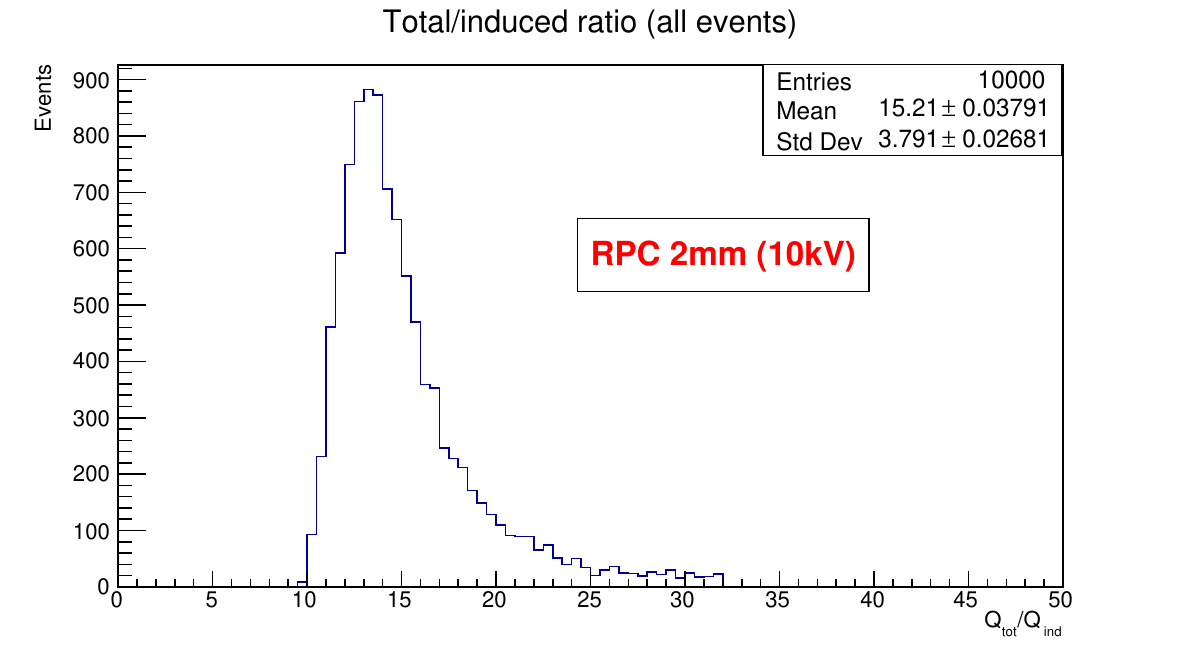}
\includegraphics[width=0.49\textwidth]{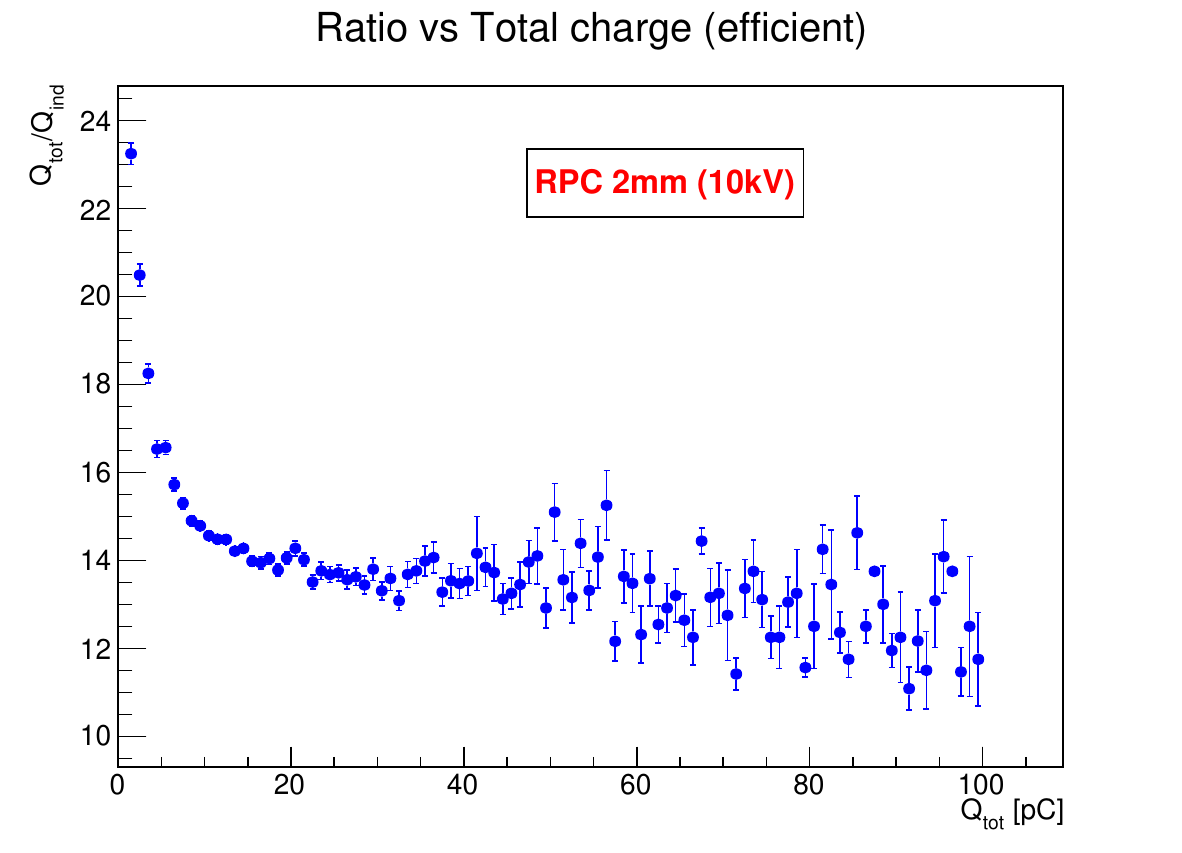}
    \caption{Distribution of the ratio between Total charge and induced charge (left) and distribution of this ratio vs the total produced charge (right), for the 2 mm gas gap RPC with standard gas mixture and operated at 10 kV.}
    \label{fig:ratioDistributionsRPC2mm}
\end{figure}

We applied the full simulation chain to the case of a planar RPC with a 
2~mm gas gap, 1.8~mm thick electrodes, and the standard gas mixture, 
operated at a voltage of 10~kV. 
Figure~\ref{fig:chargeDistributionsRPC2mm} shows the resulting distribution of the total avalanche charge 
(left panel) and of the induced charge (right panel) obtained over the simulated events.  
In Figure~\ref{fig:ratioDistributionsRPC2mm}, the distribution of the ratio between total and induced charge is shown
(left plot), together with the two-dimensional correlation plot of this ratio as a function of 
the total charge (right). These results allow one to quantify the spread of the avalanche 
size and to investigate how the fraction of induced charge varies with the avalanche magnitude.

From the simulated distributions it is found that the \emph{average} total charge is about 
$ \langle Q_{\mathrm{tot}} \rangle \approx 16~\mathrm{pC} $, while the \emph{average} induced charge is 
$ \langle Q_{\mathrm{ind}} \rangle \approx 1.25~\mathrm{pC} $.  
On average, the ratio $ Q_{\mathrm{tot}} / Q_{\mathrm{ind}} $ is about  15, 
with a clear tendency to increase for events producing small total charges.  
To estimate the detection efficiency, we define a charge threshold of 
$ Q_{\mathrm{ind,th}} = 20~\mathrm{fC} $ and consider as \emph{efficient} those events with 
induced charge above this threshold.  
With this criterion, the simulated efficiency for this configuration is above $ 98\% $.

\subsection{Simulation results for RPCs with different gas gaps as a function of the applied voltage}
\label{RPCvsHV}

Three RPC configurations with gas gaps of 2 mm, 1 mm, and 0.5 mm, electrodes of 1.8 mm thickness, and standard gas mixture have been simulated. For each configuration, 5000 events were generated at different applied voltages.  

\begin{figure}[h!]
    \centering  
    \includegraphics[width=0.48\textwidth]{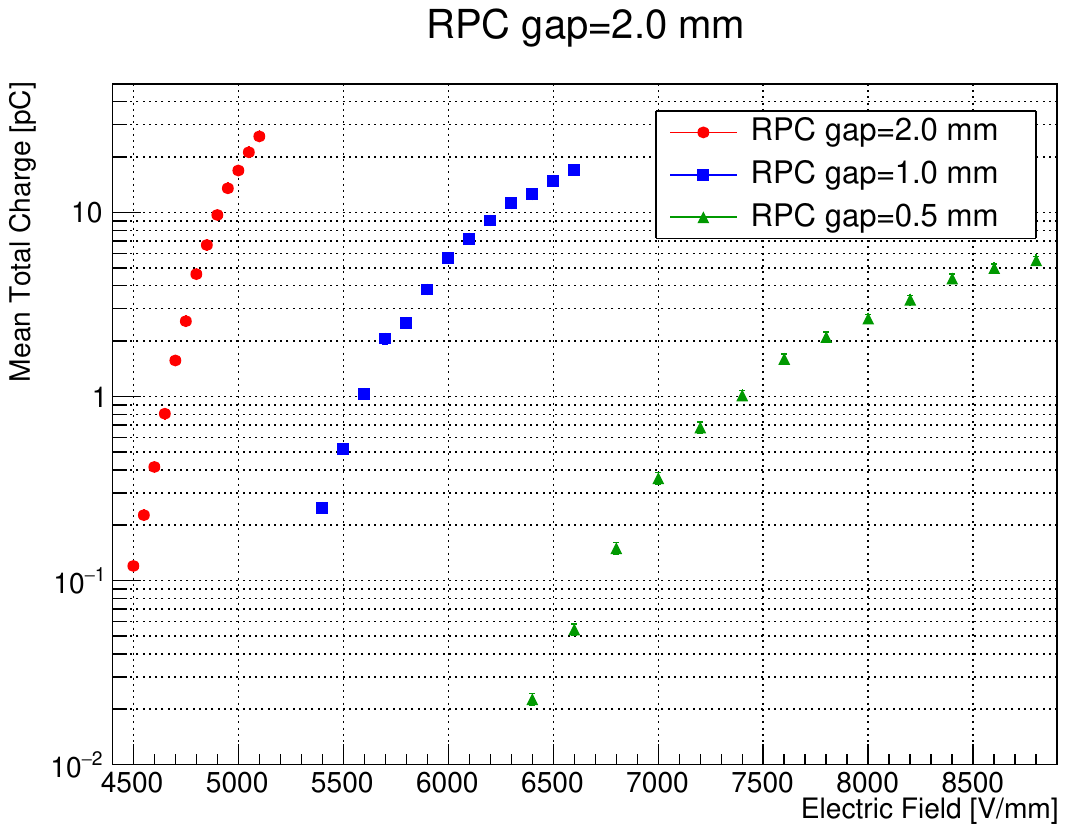}
    \includegraphics[width=0.48\textwidth]{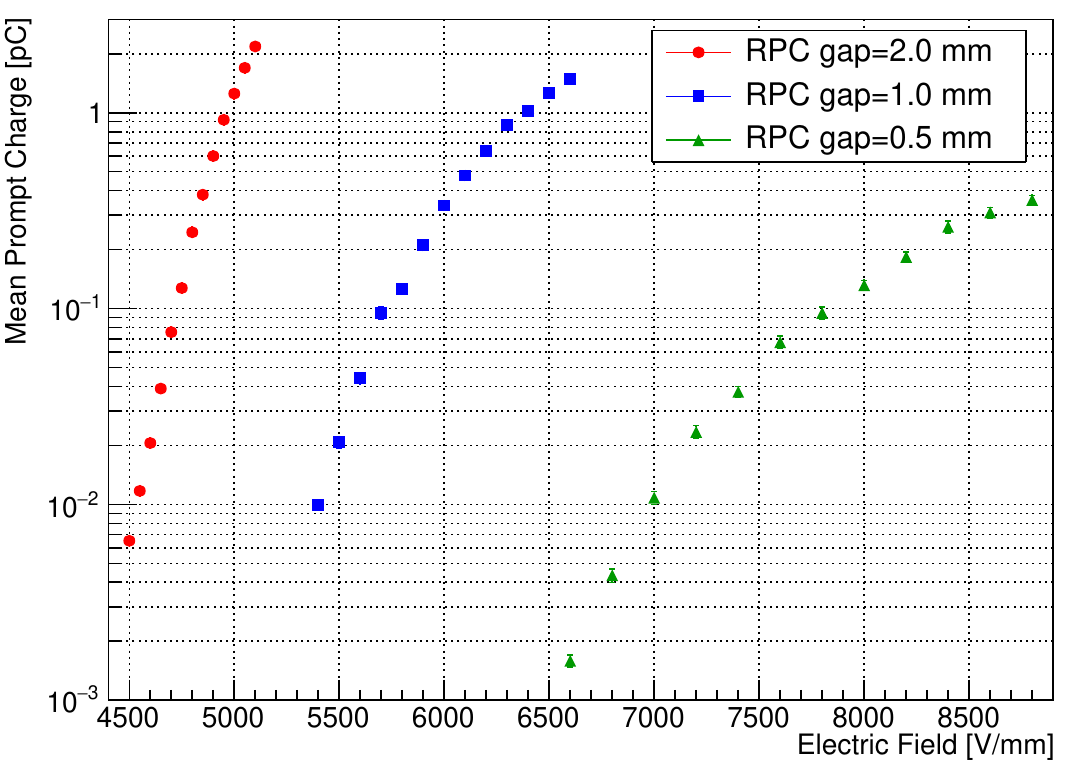}
    \caption{Average Total charge (left) and average induced charge (right) vs Electric field for 2, 1 and 0.5 mm gas gap RPCs operated with standard gas mixture.}
    \label{fig:chargeVsEfield_RPC}
\end{figure}

\begin{figure}[h!]
    \centering  
    \includegraphics[width=0.48\textwidth]{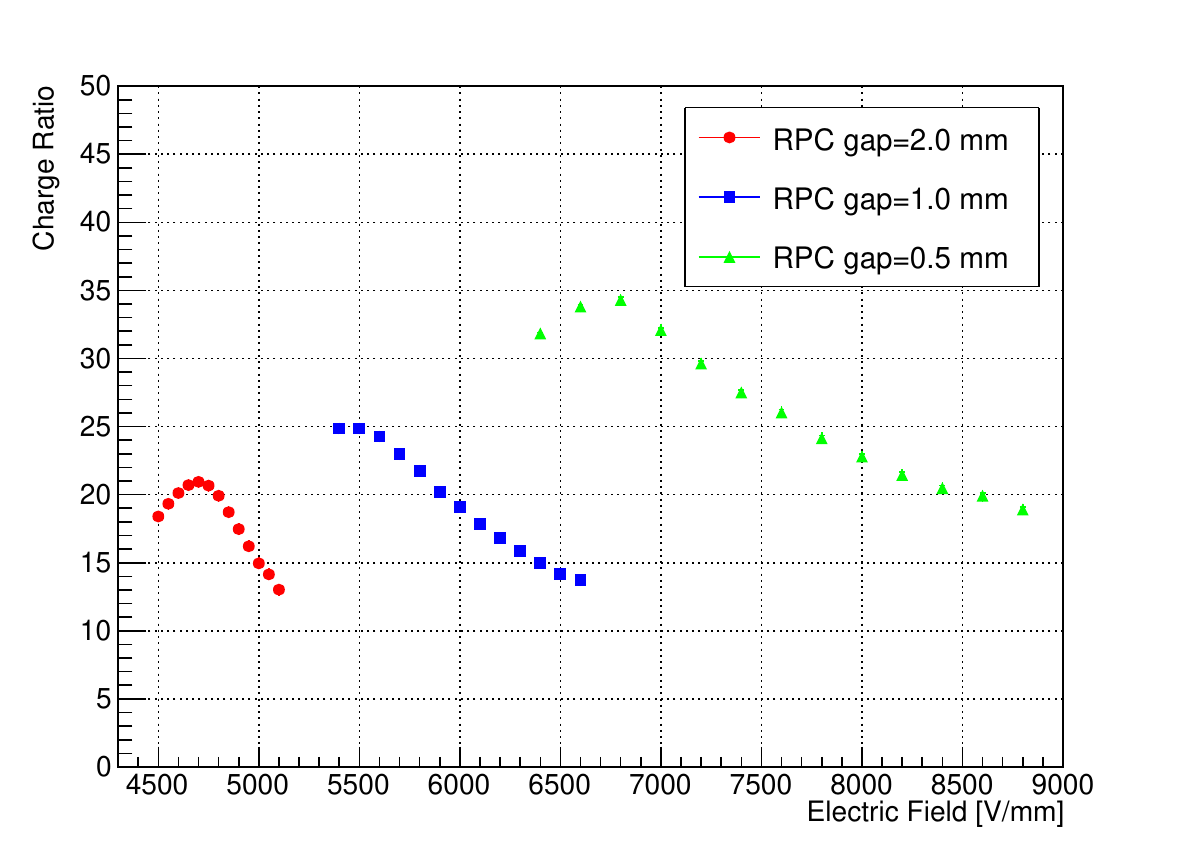}
    \includegraphics[width=0.48\textwidth]{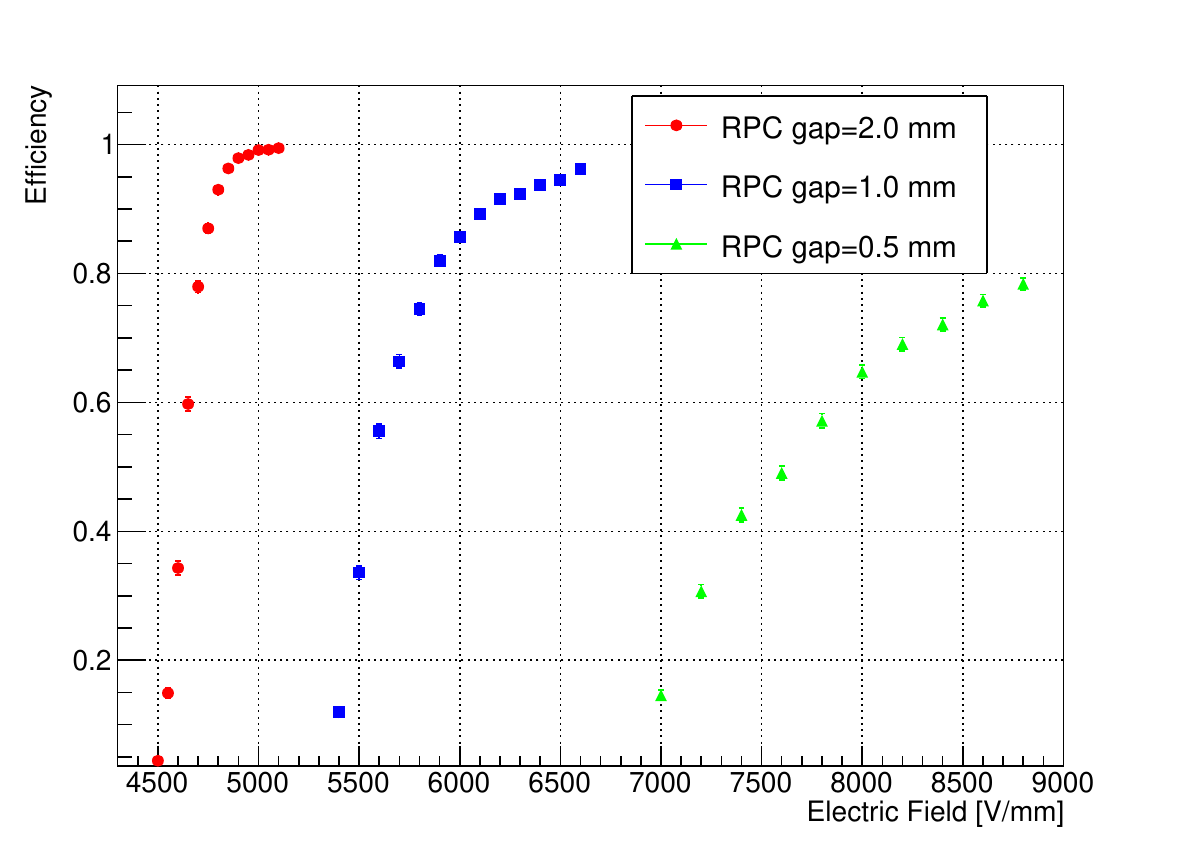}
    \caption{Ratio of Total over induced charge (left) and efficiency (right) vs Electric field for 2, 1 and 0.5 mm gas gap RPCs operated with standard gas mixture. Efficiency curve corresponds to an induced charge threshold of 20 fC. Note that the ratio is evaluated only for efficient events.} 
    \label{fig:RatioEffVsEfield_RPC}
\end{figure}

Figure \ref{fig:chargeVsEfield_RPC} shows the mean total charge (left) and mean induced charge (right) for all events, as a function of the electric field for the three different RPC layout with different gas gap. Figure \ref{fig:RatioEffVsEfield_RPC} shows the ratio of total to induced charge for efficient events (fraction of events with induced charge above 20 fC), and the detection efficiency as a function of the electric field. Each curve corresponds to a different gas gap configuration.

The simulation results reproduce very well the experimental data, as reported for example in \cite{Santonico2016}.  
For the 2 mm and 1 mm gas gap, the simulated curve matches all experimental points quite well. For the 0.5 mm the efficiency is generally about 10 \% lower than experimental results and the simulated charges are in general a factor 1.5-2 lower than the experimental data around the working point with larger discrepancies only at very low efficiency values.
Part of the discrepancy at lower voltages can be explained by the fact that we are considering tracks orthogonal to the gas gap, while, taking into account the cosmic angular distribution, an enhancement of the number of primary clusters generated and, as a consequence, the efficiency and the charge generated, can be expected.
Overall, the agreement with experimental data seems extremely encouraging despite the simplified modeling of the space-charge effect implemented.

\section{Application of the full avalanche Simulation results to the RCC geometry}
\label{RCCFullSim}

After having validated the simulation tool with the RPC geometry, we applied the full simulation framework to the case of a Resistive Cylinder Chamber (RCC) with 
an inner radius $r_i = 2~\mathrm{mm}$ and an outer radius $r_o = 3~\mathrm{mm}$. 

\begin{figure}[h!]
  \centering
  % Placeholder for the two plots side by side
  \includegraphics[width=0.48\textwidth]{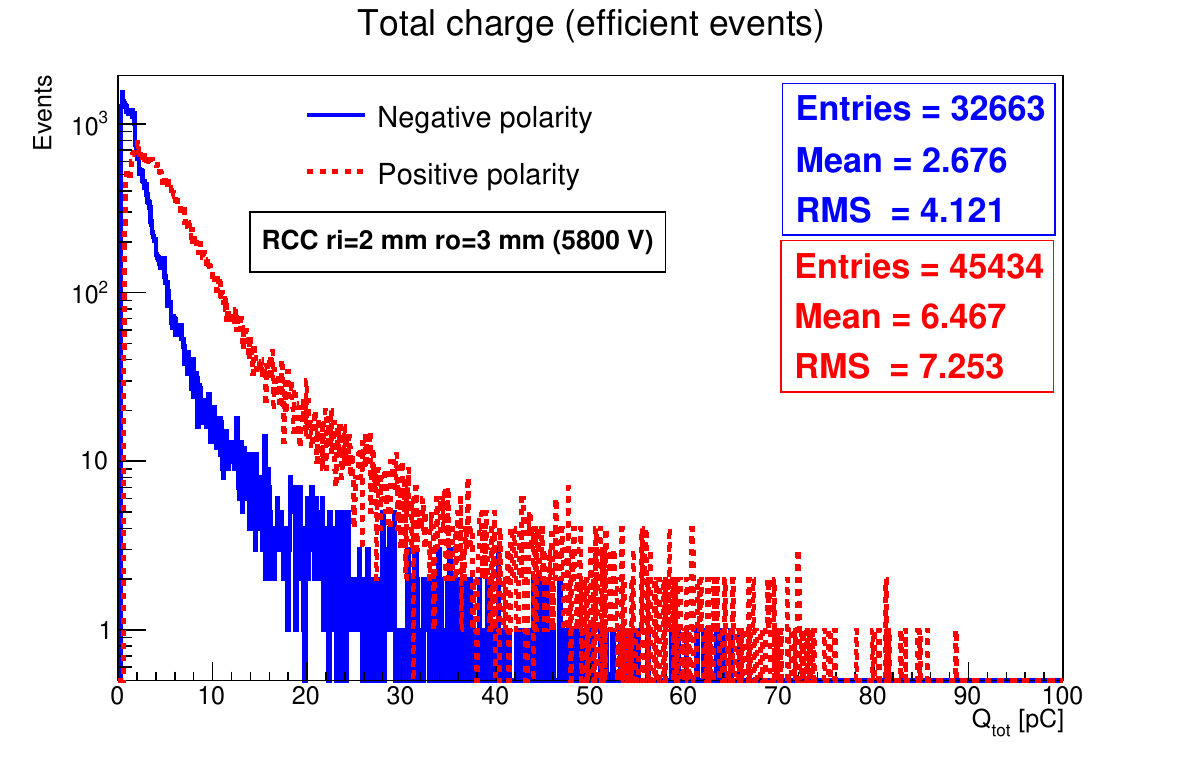}
  \includegraphics[width=0.48\textwidth]{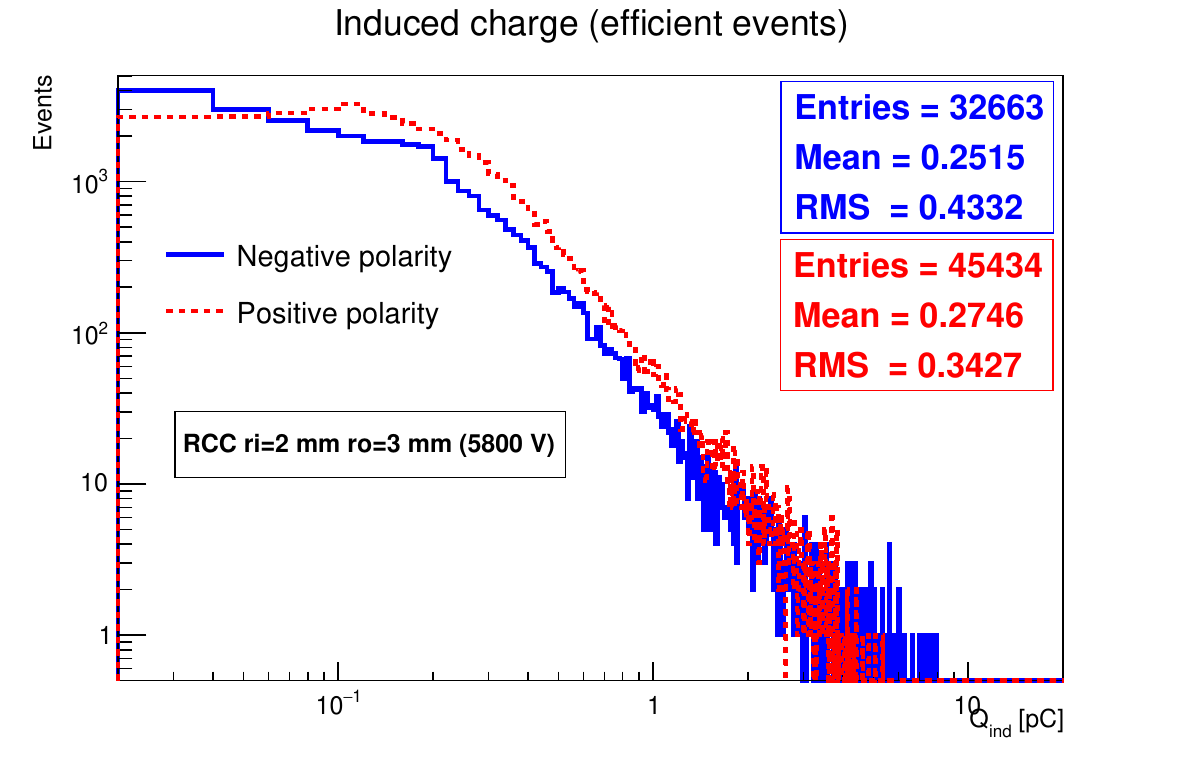}
  \caption{Simulated distributions of total charge (left) and induced charge (right) 
  for the RCC operated at 5800~V with positive and negative polarity. Only efficient events are included in the plot (induced charge above 20 fC).}
  \label{fig:RCC_5800_charges}
\end{figure}

Figures~\ref{fig:RCC_5800_charges} (left) and (right) show, respectively, the distributions of the total charge 
and the induced charge obtained for the RCC operated at a working voltage of 5800~V, for both positive and 
negative polarity configurations. 

\begin{figure}[h!]
  \centering
  \includegraphics[width=0.55\textwidth]{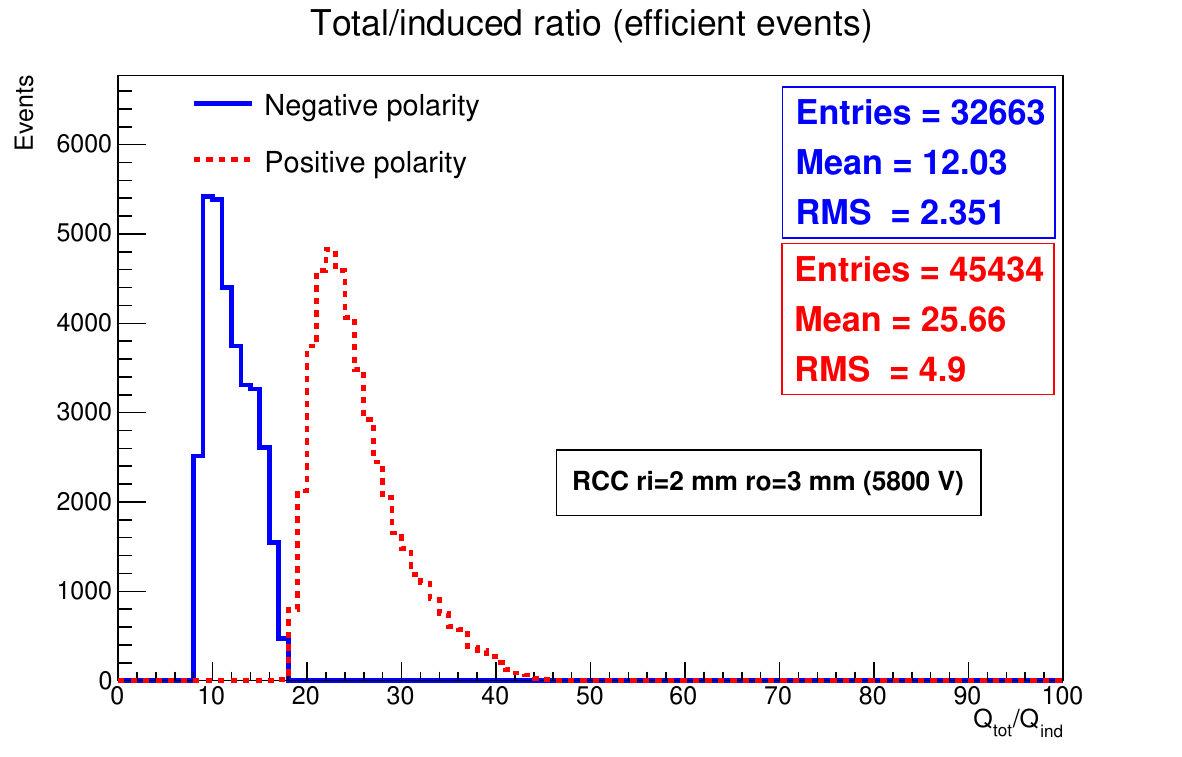}
  \caption{Distribution of the ratio between total charge and induced charge for the RCC operated at 5800~V, 
  for both positive and negative polarity. Only efficient events are included in the plot (induced charge above 20 fC).}
  \label{fig:RCC_5800_ratio}
\end{figure}

The corresponding distribution of the ratio between total and induced charge is reported in 
Fig.~\ref{fig:RCC_5800_ratio}, where again the comparison between positive and negative polarities is provided. 

\begin{figure}[h!]
  \centering
  % Placeholder for the two plots side by side
  \includegraphics[width=0.48\textwidth]{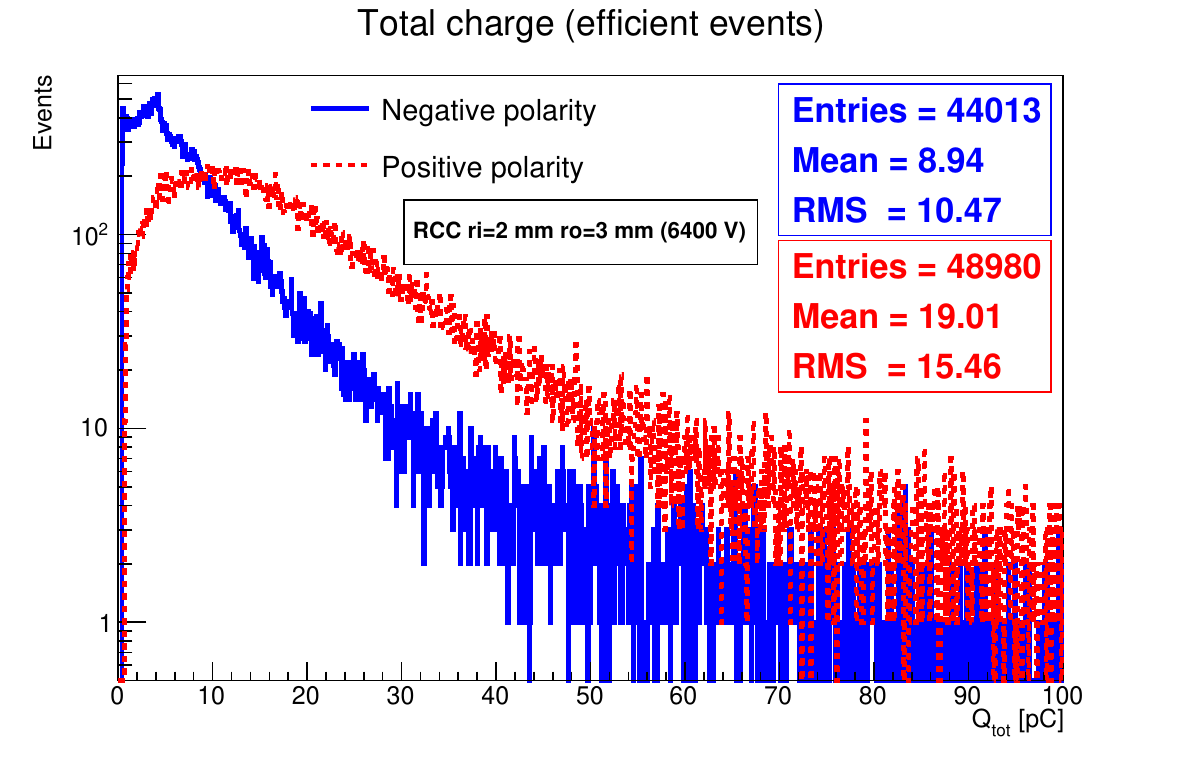}
  \includegraphics[width=0.48\textwidth]{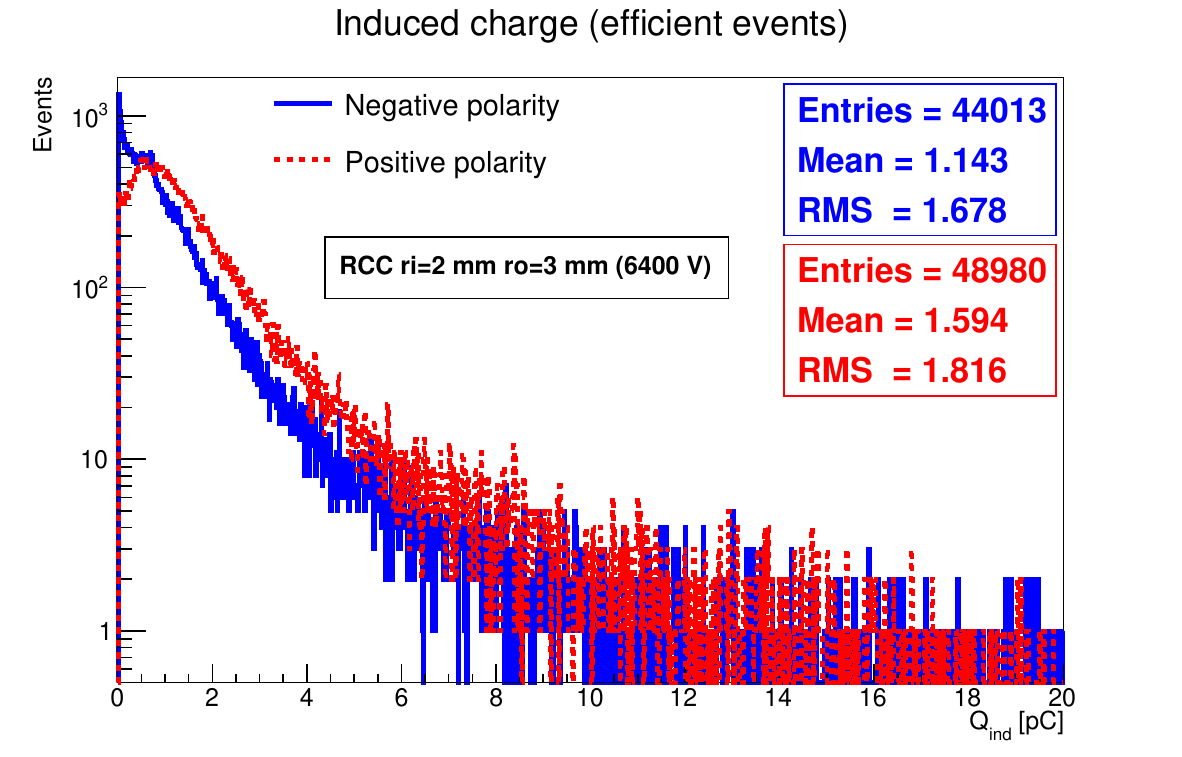}
  \caption{Simulated distributions of total charge (left) and induced charge (right) 
  for the RCC operated at 6400~V with positive and negative polarity. Only efficiency events are included in the plot (induced charge above 20 fC).}
  \label{fig:RCC_6400_charges}
\end{figure}

\begin{figure}[h!]
  \centering
  \includegraphics[width=0.55\textwidth]{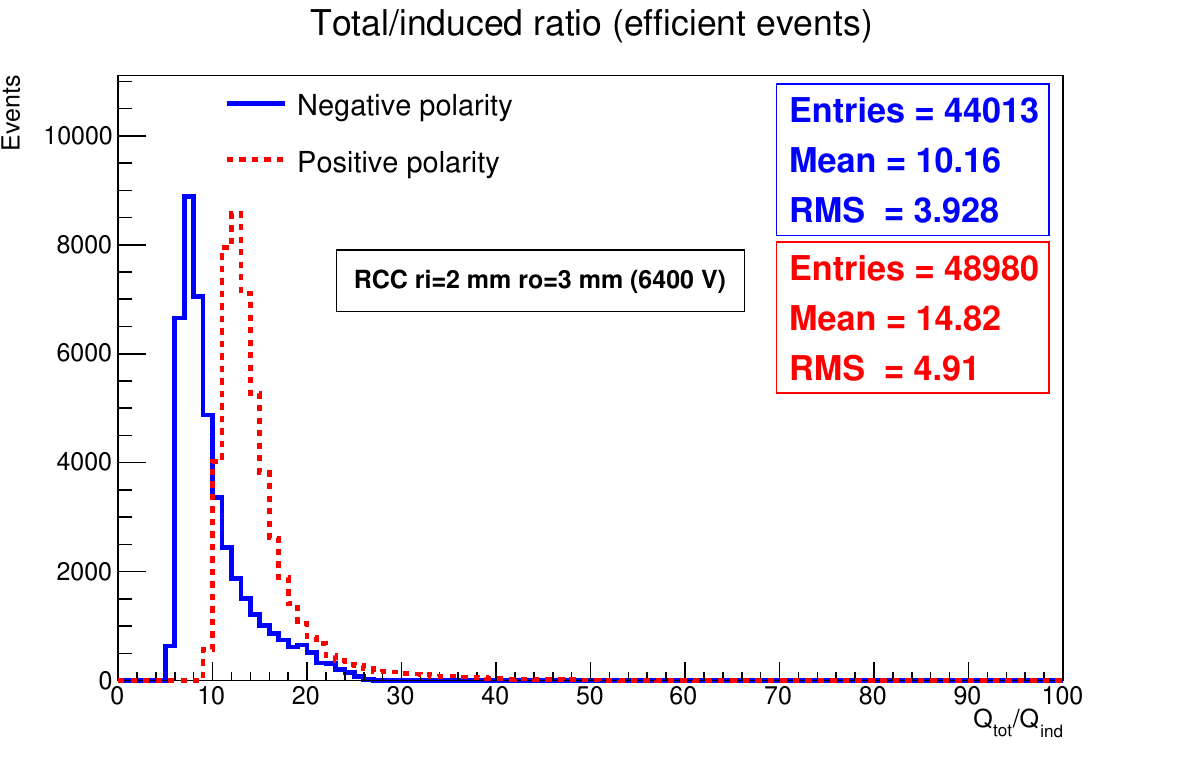}
  \caption{Distribution of the ratio between total charge and induced charge for the RCC operated at 6400~V, 
  for both positive and negative polarity.}
  \label{fig:RCC_6400_ratio}
\end{figure}

The same simulation procedure was repeated at a higher working voltage of 6400~V. The results, showing the 
distributions of total and induced charge, as well as the ratio between them, are displayed in 
Figs.~\ref{fig:RCC_6400_charges} and \ref{fig:RCC_6400_ratio}. 

 From these plots we can extract that, at 5800~V, the average total charge is about 6.5~pC for positive polarity and about 2.7~pC for negative polarity, 
while the corresponding average induced charges are 0.25~pC and 0.27~pC, respectively. 
At 6400~V, the average total charge rises to approximately 19~pC for positive polarity and about 9~pC for negative polarity, 
with corresponding induced charges of 1.6~pC and 1.1~pC. 

However, when examining the number of entries of the plots restricted to efficient events 
(i.e.\ with $Q_{\mathrm{ind}}>20$~fC), over a total of 50,000 events generated, the detection efficiency can be extracted. 
At 5800~V, the efficiency is about 91\% for positive polarity and 65\% for negative polarity. 
At 6400~V, the efficiencies increase to about 98\% and 88\%, respectively. 

In conclusion, although the ratio $Q_{\mathrm{tot}}/Q_{\mathrm{ind}}$ tends to favor the negative polarity, 
the positive polarity configuration provides significantly higher efficiencies at the same operating voltage, 
in line with the qualitative expectations discussed at the end of section \ref{InducedCharge_realisticConfigurations}

\subsection{Simulation results for RCCs with different gas gaps as a function of applied voltage}

As for the RPC in section \ref{RPCvsHV}, we simulated the RCC response as a function of the applied voltage for different gas gaps. Fixing the inner radius of the RCC $r_i$ = 2 mm, we studied the total charge, the induced charge, the charge ratio and the efficiency for gas gaps: 1 mm and 0.5 mm and for both positive and negative polarities. The electrode thickness is always fixed at 1 mm. For each configuration, 10.000 events were generated at different applied voltages.  

\begin{figure}[h!]
    \centering  
    \includegraphics[width=0.48\textwidth]{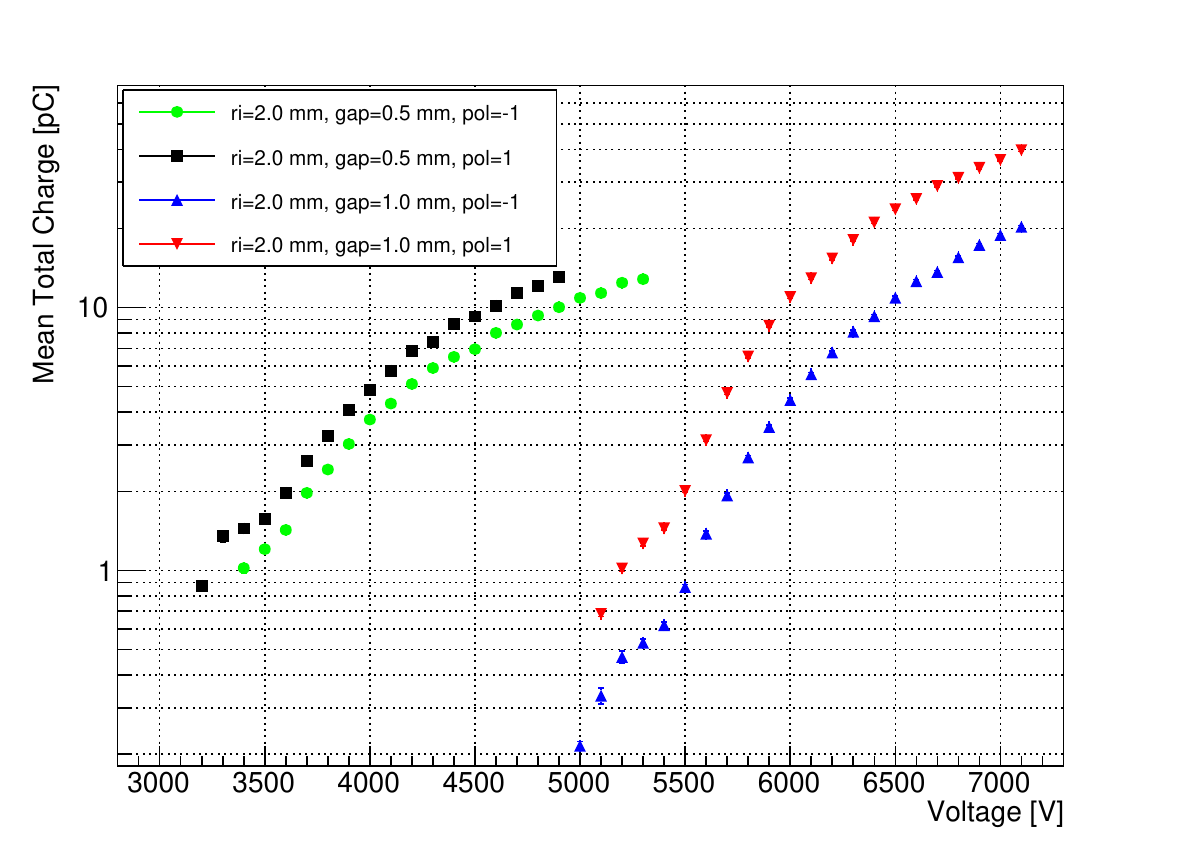}
    \includegraphics[width=0.48\textwidth]{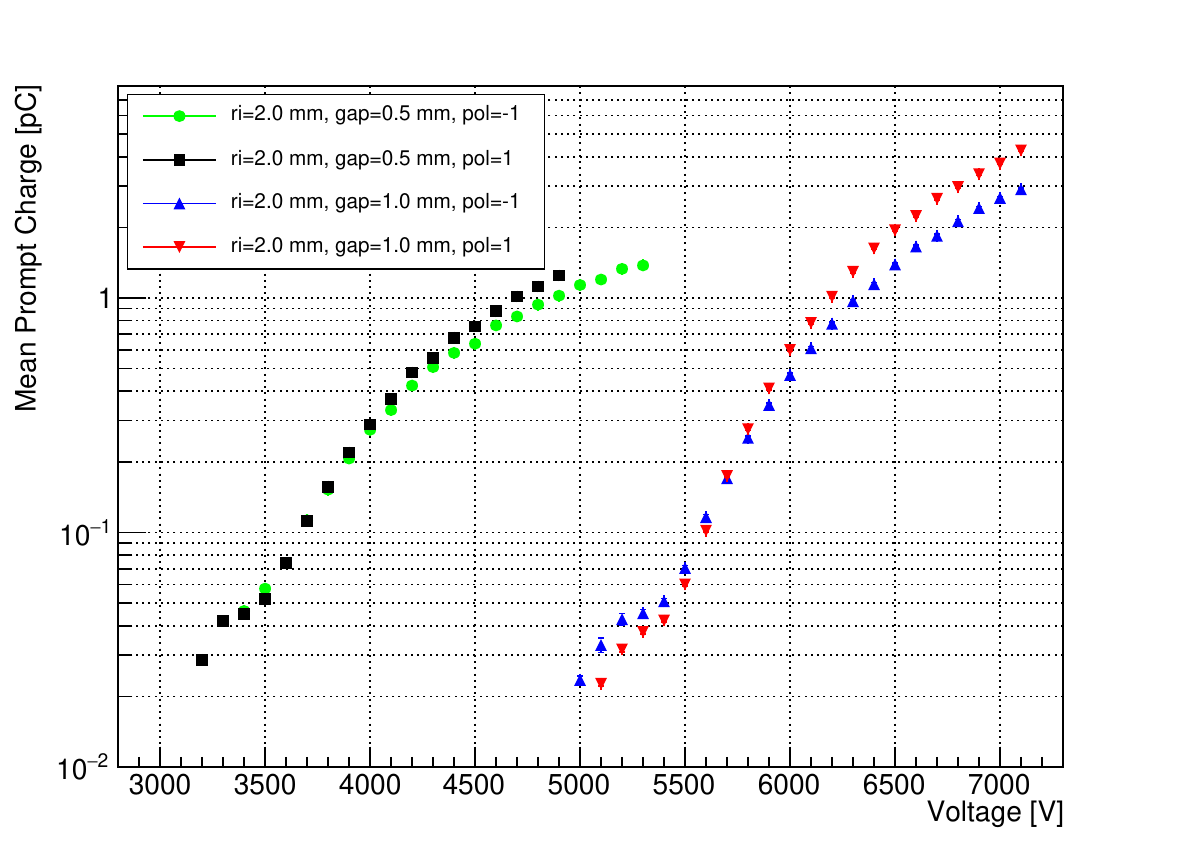}
    \caption{Average Total charge (left) and average induced charge (right) vs Electric field for 1 and 0.5 mm gas gap RCCs with inner radius $r_i$= 2 mm and operated with standard gas mixture in both positive and negative polarity. Only efficient events have been included in the average calculation (induced charge $>$ 20 fC)}
    \label{fig:chargeVsEfield_RCC}
\end{figure}

\begin{figure}[h!]
    \centering  
    \includegraphics[width=0.48\textwidth]{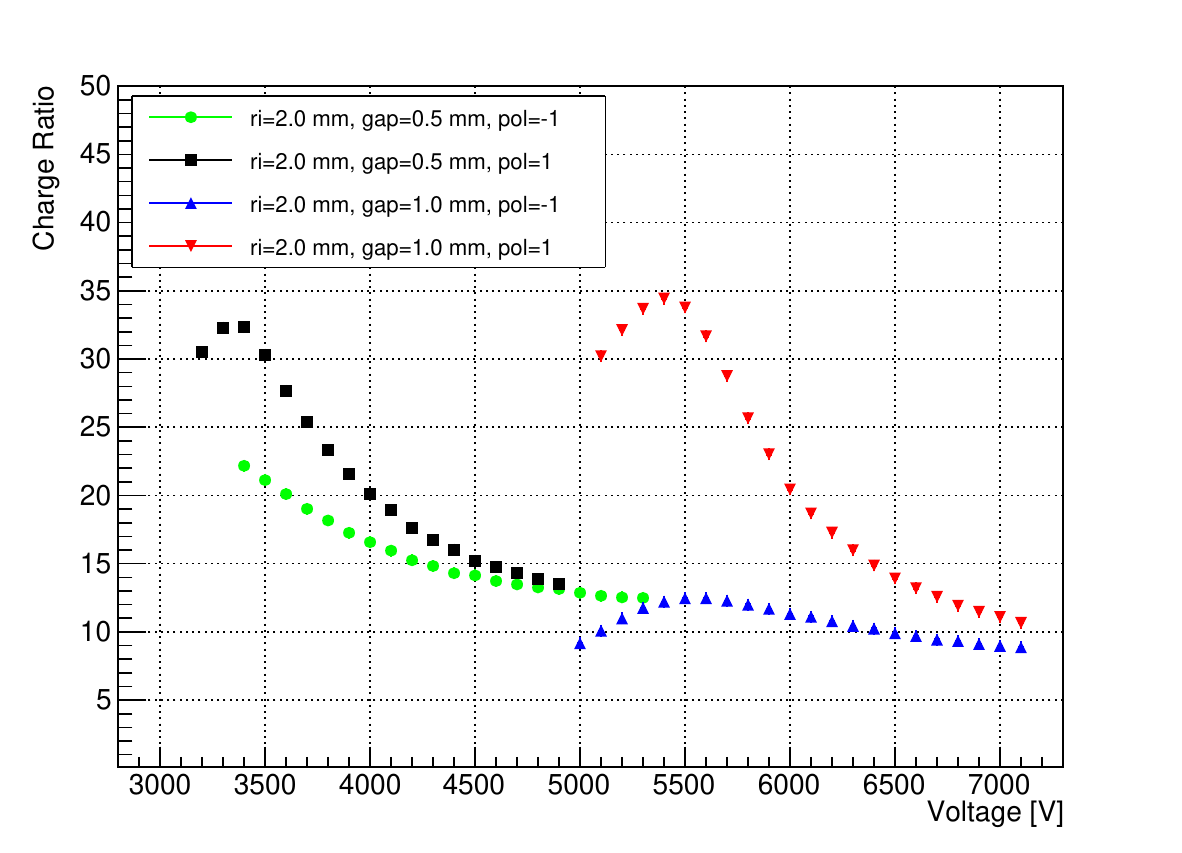}
    \includegraphics[width=0.48\textwidth]{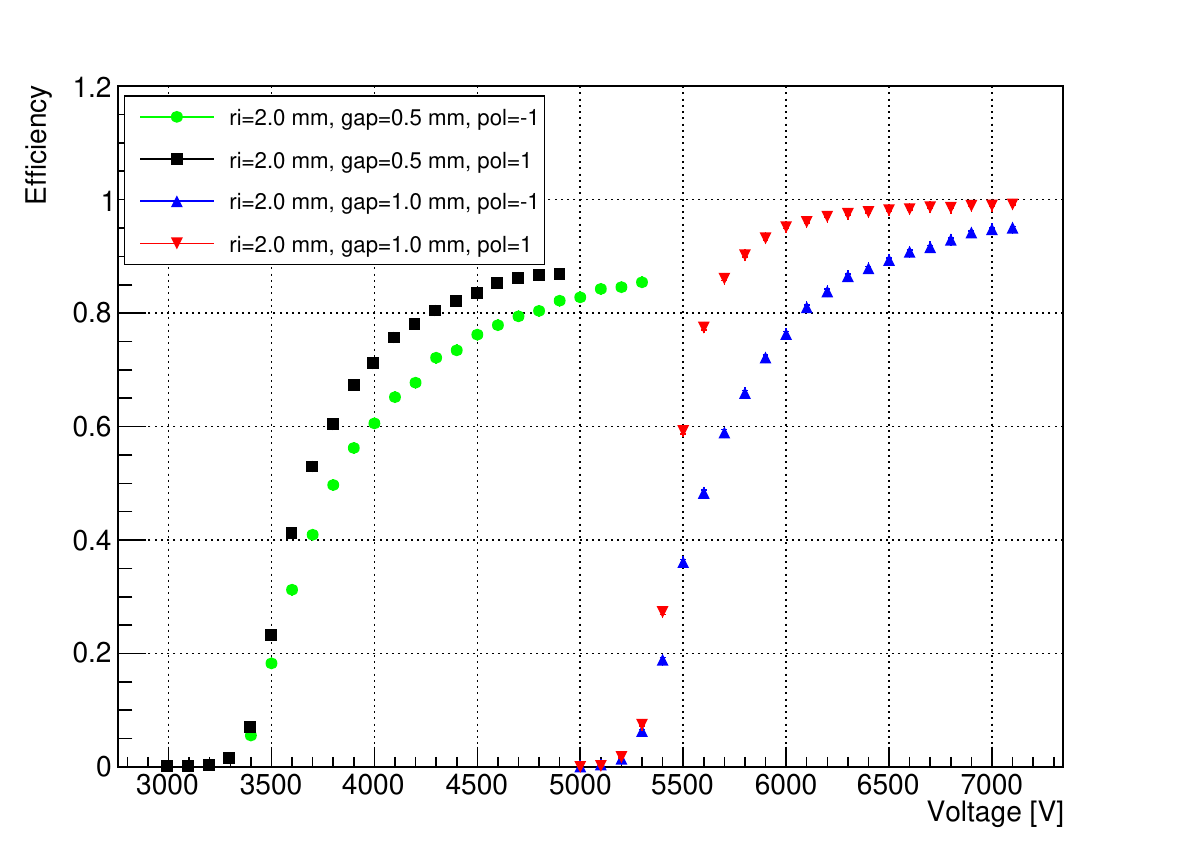}    
    \caption{Ratio of total over induced charge (left) and efficiency (right) vs Electric field for 1 and 0.5 mm gas gap RCCs with inner radius $r_i$= 2 mm and operated with standard gas mixture in both positive and negative polarity. Efficiency curve corresponds to an induced charge threshold of 20 fC. as in the previous case the ratio is evaluated only for efficient events.} 
    \label{fig:RatioEffVsEfield_RCC}
\end{figure}

Figure \ref{fig:chargeVsEfield_RCC} shows the mean total charge (left) and mean induced charge (right), as a function of the electric field for efficient events (fraction of events with induced charge above 20 fC), and for the two RCC layouts with different gas gap and different polarities. Figure \ref{fig:RatioEffVsEfield_RCC} shows the ratio of total to induced charge for efficient events, and the detection efficiency as a function of the electric field. Each curve corresponds to a different gas gap/polarity configuration.

\begin{figure}[h!]
    \centering  
    \includegraphics[width=0.48\textwidth]{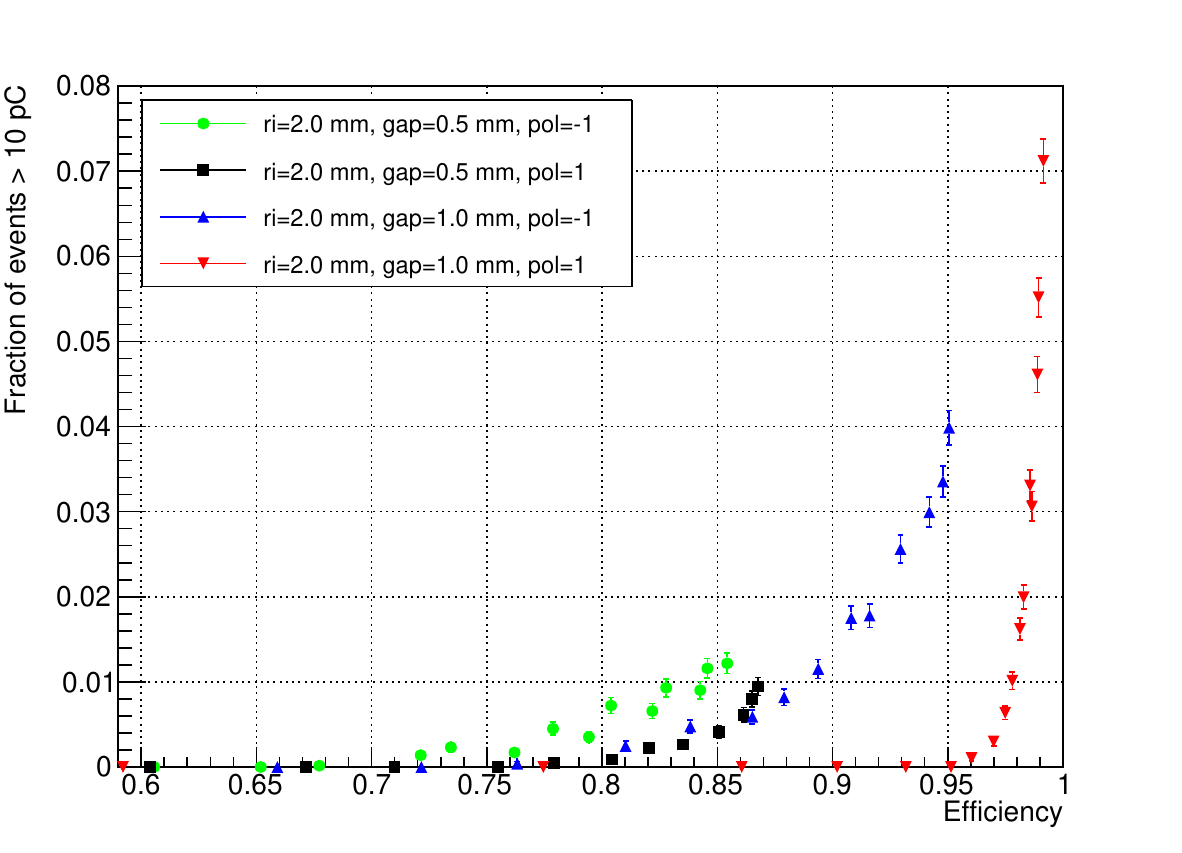}
    \includegraphics[width=0.48\textwidth]{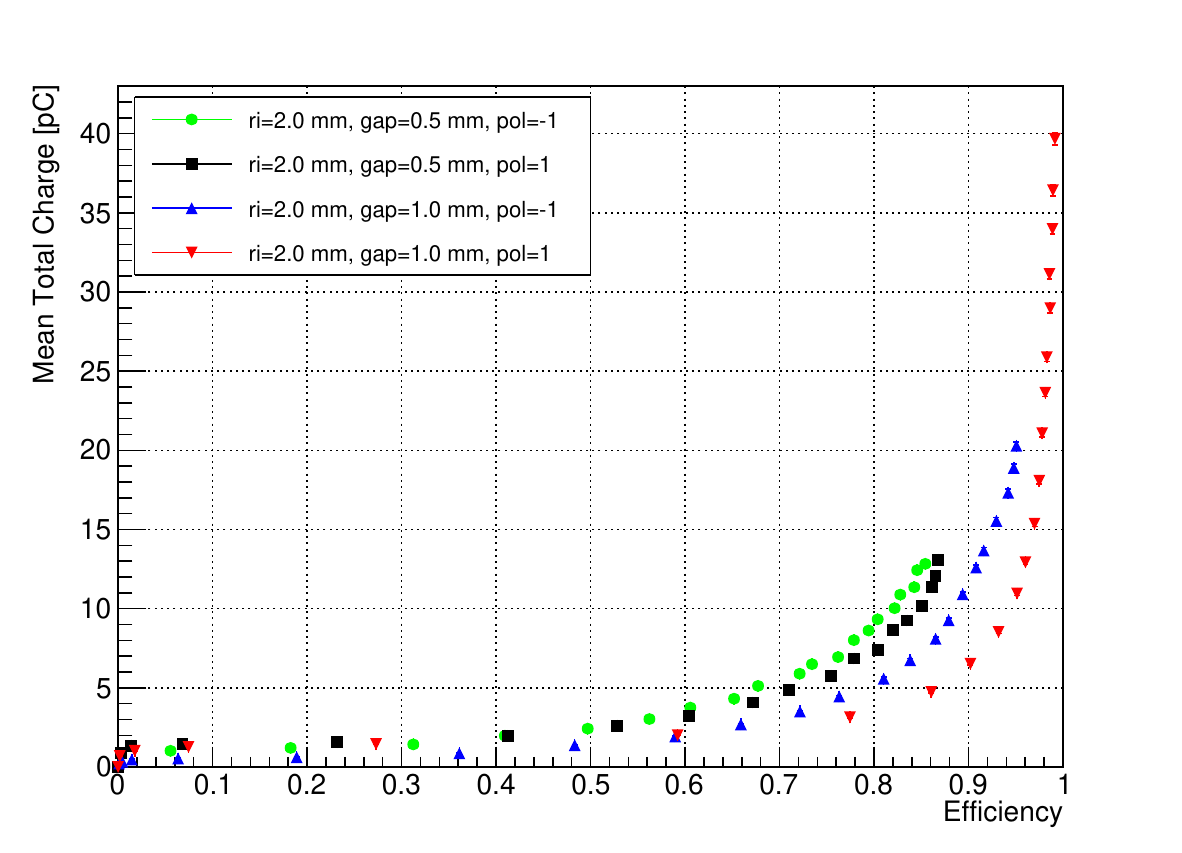}
    \caption{Fraction of events with induced charge above 10 pC (left) and total charge (right) as a function of the detection efficiency (induced charge above 20 fC), for the two different gas gaps and the two different polarities.} 
    \label{fig:ChargesVsEff_RCC}
\end{figure}

As the different layouts reach higher efficiencies at different voltage values, it is interesting to check the fraction of events with large produced charges and the average total charge as a function of the detection efficiency. In figure \ref{fig:ChargesVsEff_RCC} we plot the fraction of events with induced charge above 10 pC and the average total charge as a function of the efficiency for the different layouts.


\subsubsection{Discussion}

Figures from ~\ref{fig:RCC_5800_charges} to \ref{fig:ChargesVsEff_RCC} clearly demonstrate the asymmetric behavior of the RCC when operated in the two opposite polarities. At equal operating voltage, both the average total charge and the average induced charge are larger in positive polarity compared to negative polarity. However, while the total charge shows a significant increase, the induced charge is only slightly larger in positive polarity than in negative polarity. This feature is most evident in Figures~\ref{fig:RCC_5800_ratio}, ~\ref{fig:RCC_6400_ratio}, and ~\ref{fig:RatioEffVsEfield_RCC}  (left panel), which display the ratio $Q_{\rm tot}/Q_{\rm induced}$.  

On the other hand, Figures~\ref{fig:RCC_5800_charges} and ~\ref{fig:RCC_6400_charges} highlight that in positive polarity the charge distribution deviates from the nearly exponential trend that characterizes the negative polarity. This deviation results in a clear benefit in terms of detection efficiency, once a fixed threshold is applied by the front-end electronics. Figure~\ref{fig:ChargesVsEff_RCC} provides even more compelling evidence: at equal efficiency, both the average total charge and the fraction of events with induced charge above 10~pC are much smaller in positive polarity compared to negative polarity.  

% parte da nascondere per ora
%The case of the 2~mm gas gap operated in positive polarity is particularly striking. It shows a plateau curve with extremely steep rise compared to the other configurations, achieving high detection efficiency at very low total charges and with practically no events having induced charge above 10~pC.  

As a final remark, it is important to stress that all the results presented here refer to tracks generated from the center of the cylindrical region and crossing the RCC radially. In our simulations, primary clusters were therefore generated only along the thickness of the gas gap. Cases where the track inclination increases the number of primary clusters, or where a secant trajectory crosses the cylindrical gas ring producing two avalanches in distinct regions, have not been considered. Such scenarios could slightly modify the conclusions described above and will be the subject of future studies.

\section{Conclusions}
\label{conclusions}

In this work we have developed a standalone simulation program, designed to be both extremely flexible and computationally efficient, aimed at quantitatively analyzing the operation of the Resistive Cylinder Chamber (RCC) as a function of its geometrical parameters. The code simulates avalanche development in both RPC and RCC geometries, makes use of the Townsend coefficient as a function of the electric field across the gas gap, and implements the effect of space charge by approximating the ionic distribution as a series of disks. The radius of each disk is linked to the transverse diffusion of the avalanche and the disks are distributed along the avalanche path. Moreover, the code computes the induced charge on external electrodes by explicitly taking into account the weighting field for both planar and cylindrical configurations.  

The program has first been validated on planar RPC geometries, where the simulation results were compared to existing experimental measurements. An excellent agreement was found, both in terms of the total and induced charges as well as the detection efficiency for gas gaps of 2 and 1 mm (the agreement is worse  for the 0.5 mm gas gap), once an appropriate threshold on the induced charge was applied. The simulation framework was then extended to the cylindrical geometry, producing the essential figures needed to characterize the novel configuration: the electric field distribution in the gas gap for the two polarities of operation, including the role of space charge; the RCC gain as a function of the avalanche starting point inside the gap; and the expected ratio between total and induced charge. These results highlight the impact of geometry and polarity on the avalanche dynamics and provide valuable insight into the operation of the new detector concept.  

All results presented were obtained assuming the standard gas mixture, taking into account both primary ionization along the track and secondary emission, as well as the field dependence of the Townsend coefficient.  

Naturally, the present simulation includes a number of simplifications. The main limitations are the absence of statistical fluctuations on the avalanche gain, the simplified description of space charge as static disks along the avalanche path, and the lack of correlations between avalanches seeded by different primary clusters (which are presently treated as fully independent). Furthermore, results have been  shown only for tracks strictly perpendicular to the electrodes in planar RPCs, or strictly radial in RCCs; secant tracks in RCC geometry, which could generate two avalanches at different points of the cylindrical gas ring, and different path lengths in the gas gap, have not been considered at this stage.  

Some of these limitations can be overcome in future developments of the code. Nevertheless, the tool already provides a powerful guide to the experimental studies that will be conducted inside the TANGO\_RD project or elsewhere, and a solid basis to assess both the potential and the limits of the novel RCC detector concept.

\appendix
\section*{Appendix}
\addcontentsline{toc}{section}{Appendix} % opzionale per inserirlo nell'indice

\section{Electric Field of a Charged Disk at distance s}

Consider a disk of radius \( a \) uniformly charged with surface charge density \( \sigma \). 
The electric field at a point on the axis of the disk at distance \( s \) from its center is obtained by integrating the contribution of concentric infinitesimal rings of charge. 
For an infinitesimal ring of radius \( r \) and charge element 
\[
dq = \sigma \cdot 2\pi r \, dr,
\]
the axial electric field component is
\[
dE = \frac{1}{4 \pi \epsilon_0} \frac{dq \cdot s}{\left(s^2 + r^2\right)^{3/2}} 
= \frac{1}{4 \pi \epsilon_0} \frac{\sigma \, 2\pi r \, dr \cdot s}{\left(s^2 + r^2\right)^{3/2}}.
\]
Integrating \( r \) from \( 0 \) to \( a \) yields the total axial field:
\begin{equation}
E(s) = \frac{\sigma}{2 \epsilon_0} \left(1 - \frac{s}{\sqrt{s^2 + a^2}}\right),
\label{SpaceChargeDisk}
\end{equation}
where \( \sigma \) is the surface charge density, \( s \) is the distance from the ion disk (assumed frozen during the avalanche), and \( a \) is the disk radius.

\section{Stepwise Calculation of Induced Charge}
In the stepwise calculation of the induced charge, we assume that the electric field \( E \) and the effective Townsend coefficient \( \alpha_{\rm eff} \) do not vary significantly within each individual step of length \(\Delta r\). This allows us to approximate the electron multiplication \( n_e \) over the step using the exponential gain formula:

\begin{equation}
n(r + \Delta r) = n(r) \, e^{\alpha_{\rm eff} \Delta r}.
\end{equation}

Correspondingly, the induced charge contribution \(\Delta Q_{\rm ind}\) from the charge generated in the step can be expressed as the integral of the charge multiplied by the weighting field \( E_w \):

\begin{equation}
\Delta Q_{\rm ind} = e \, E_w(r_{\rm mid}) \int_0^{\Delta r} n(r) \, e^{\alpha_{\rm eff} x} \, dx = e \, E_w(r_{\rm mid}) \, n(r) \frac{e^{\alpha_{\rm eff} \Delta r} - 1}{\alpha_{\rm eff}},
\end{equation}

where \( r_{\rm mid} \) is the midpoint of the step. In the limit \(\alpha_{\rm eff} \times \Delta r\ll 1\), this reduces to

\begin{equation}
\Delta Q_{\rm ind} \approx e \, E_w(r_{\rm mid}) \, n(r) \, \Delta r.
\end{equation}

By dividing the total path into \(N\) steps (typically between 100 and 1000), we treat \(E\) and \(\alpha_{\rm eff}\) as constant within each step and compute the cumulative induced charge as

\begin{equation}
Q_{\rm ind} = \sum_{i=1}^N \Delta Q_{\rm ind}^{(i)}.
\end{equation}

With this discretization, the approximation error due to neglecting the variation of \( n_e \) within each step is estimated to be less than 10\%, which is acceptable for the current level of modeling. Reducing the step size further can improve accuracy if needed.

This approach thus provides a computationally efficient and sufficiently accurate method to evaluate the total induced charge along the avalanche path, including the effect of space charge when enabled.

\subsection{Evaluation of the Weighting Field for Planar and Cylindrical Geometries}

The induced charge on the electrodes in gaseous detectors such as Resistive Plate Chambers (RPCs) and Resistive Cylindrical Chambers (RCCs) can be calculated using the Shockley-Ramo theorem, which relates the instantaneous induced current to the movement of charges within the detector via the \emph{weighting field} \cite{Ramo1939}.

The weighting field \(\vec{E}_w\), used in equation \ref{ramostokes}, is defined as the electric field configuration in the detector when the electrode of interest is set to unit potential (1 V) and all other electrodes are grounded, \emph{neglecting} the presence of space charge.

In RPCs and RCCs, the electrodes are typically made of semi-insulating materials with relative permittivity \(\epsilon_r\) and thickness \(d\) on the order of millimeters. The weighting field depends on the geometry and dielectric properties of the system and must be evaluated accordingly.
\subsubsection*{Planar Geometry}

Consider a planar RPC with two parallel electrodes separated by a gas gap of thickness \(g\), and electrode thickness \(d\). We define a 1D coordinate \(x\) perpendicular to the electrodes, with \(x=0\) at the electrode of interest and \(x=g\) at the opposite electrode.

The weighting potential \(\phi_w(x)\) satisfies Laplace’s equation in one dimension:
\[
\frac{d^2 \phi_w}{dx^2} = 0,
\]
with boundary conditions:
\[
\phi_w(0) = 1, \quad \phi_w(g) = 0.
\]

However, the electrodes have finite thickness \(d\) and relative permittivity \(\epsilon_r\), which effectively extend the gap because the electric displacement field \(D = \epsilon E\) must be continuous across interfaces. Modeling the system as three dielectric slabs in series (electrode - gas gap - electrode), the total effective thickness for the weighting field becomes:
\[
g_{\text{eff}} = g + 2 \frac{d}{\epsilon_r},
\]
where the electrode thickness is scaled by \(1/\epsilon_r\) because of the dielectric permittivity.

The linear potential drop across the effective gap leads to a constant weighting field magnitude in the gas region:
\begin{equation}
E_w = -\frac{d\phi_w}{dx} = \frac{1}{g_{\text{eff}}} = \frac{1}{g + 2 d/\epsilon_r}.
\label{eq:weight_planar_explicit}
\end{equation}

\subsubsection*{Cylindrical Geometry: Ramo Theorem Setup and Exact Weighting Field}

We consider a coaxial detector with inner electrode at radius $r_i$ and outer electrode at radius $r_o$. The gas fills $r_i<r<r_o$. Each electrode is a semi-insulating dielectric plate of thickness $d$ and relative permittivity $\varepsilon_r$. For the \emph{weighting} problem (Ramo–Shockley theorem), one electrode is set to unit potential (the \emph{readout} electrode) and the outer boundary is set to zero; no free charges are present. The goal is to obtain the radial weighting field $E_w(r)$ in the gas, which determines induced current/charge via $i(t)=\sum_k q_k \mathbf{v}_k\!\cdot\!\mathbf{E}_w(\mathbf{r}_k)$.

\paragraph{Model and strategy.}
Because of cylindrical symmetry and piecewise-homogeneous media, the weighting potential solves Laplace’s equation in three concentric annuli: inner dielectric ($r_i-d<r<r_i$), gas ($r_i<r<r_o$), outer dielectric ($r_o<r<r_o+d$). We solve $\frac{1}{r}\frac{d}{dr}\!\left(r\frac{d\phi}{dr}\right)=0$ in each region, enforce continuity of potential $\phi$ and normal displacement $D_r=\varepsilon E_r$ at the two interfaces, and apply the boundary conditions at the inner and outer external surfaces. This yields the \emph{exact} gas-region weighting field.

\paragraph{Exact three-region solution.}
In each region $k=1,2,3$ (inner dielectric, gas, outer dielectric),
\[
\phi_k(r)=A_k\ln r + B_k,\qquad E_{r,k}(r)=-\frac{d\phi_k}{dr}=-\frac{A_k}{r}.
\]
Impose:
\[
\begin{aligned}
&\phi_1(r_i-d)=1, \qquad \phi_3(r_o+d)=0,\\
&\phi_1(r_i)=\phi_2(r_i), \quad \phi_2(r_o)=\phi_3(r_o),\\
&\varepsilon\,A_1=\varepsilon_0\,A_2,\quad \varepsilon_0\,A_2=\varepsilon\,A_3 \quad (\varepsilon=\varepsilon_r\varepsilon_0).
\end{aligned}
\]
Eliminating the integration constants, one obtains a closed form for the gas-region field $E_w(r)=-d\phi_2/dr$:
\begin{equation}
E_w(r) =
\frac{1}{%
  r \, \Biggl\{
    \ln\frac{r_o}{r_i} + \frac{1}{\varepsilon_r} \Bigl(
      \ln\frac{r_i}{r_i-d} - \ln\frac{r_o}{r_o+d}
    \Bigr)
  \Biggr\}%
}, \qquad r \in (r_i,r_o).
\label{eq:Ew_exact}
\end{equation}
\paragraph{Consistency checks and usage.}
\begin{itemize}
  \item In the limit $d\!\to\!0$ (vanishing electrode thickness), \eqref{eq:Ew_exact} reduces to the familiar gas-only result $E_w(r)=1\big/\!\big[r\,\ln(r_o/r_i)\big]$.
  \item Equation \eqref{eq:Ew_exact} is the recommended expression to use in simulations whenever electrode thickness and permittivity are non-negligible.
\end{itemize}

\bibliographystyle{alpha}
\bibliography{sample}

\end{document}